\documentclass[english]{emulateapj}
\usepackage[T1]{fontenc}
\usepackage[latin9]{inputenc}
\usepackage{verbatim}
\usepackage{amssymb}
\usepackage{color}
\begin{document}

\title{Effects of Dynamical Evolution of Giant Planets on the Delivery of Atmophile Elements during Terrestrial Planet Formation}

\author{Soko Matsumura\altaffilmark{1,2}}
\author{Ramon Brasser\altaffilmark{3}}
\author{Shigeru Ida\altaffilmark{3}}

\altaffiltext{1}{School of Engineering, Physics, and Mathematics, University of Dundee, Scotland DD1 4HN}
\email{s.matsumura@dundee.ac.uk}
\altaffiltext{2}{Dundee Fellow}
\altaffiltext{3}{Earth-Life Science Institute, Tokyo Institute of Technology, Meguro-ku, Tokyo, 152-8550, Japan}

\begin{abstract}
Recent observations started revealing the compositions of protostellar discs and planets beyond the Solar System.
In this paper, we explore how the compositions of terrestrial planets are affected by dynamical evolution of giant planets.
We estimate the initial compositions of building blocks of these rocky planets by using a simple condensation model, 
and numerically study the compositions of planets formed in a few different formation models of the Solar System.
We find that the abundances of refractory and moderately volatile elements are nearly 
independent of formation models, 
and that all the models could reproduce the abundances of these elements of the Earth.
The abundances of atmophile elements, on the other hand, depend on the scattering rate of icy planetesimals 
into the inner disc as well as the mixing rate of the inner planetesimal disc.
For the classical formation model, neither of these mechanisms are efficient 
and the accretion of atmophile elements during the final assembly of terrestrial planets 
appears to be difficult.
For the Grand Tack model, both of these mechanisms are efficient, which leads to a relatively uniform 
accretion of atmophile elements in the inner disc.
It is also possible to have a ``hybrid'' scenario where the mixing is not very efficient 
but the scattering is efficient. 
The abundances of atmophile elements in this case increases with orbital radii. 
Such a scenario may occur in some of the extrasolar planetary systems which are not accompanied by giant planets or 
those without strong perturbations from giants.  
We also confirm that the Grand Tack scenario leads to the distribution of asteroid analogues 
where rocky planetesimals tend to exist interior to icy ones, and show that 
their overall compositions are consistent with S-type and C-type chondrites, respectively.   
\end{abstract}


%
\section{Introduction}\label{intro}
Recent observations of extrasolar planetary systems have shown that even similar-sized planets have 
a wide range of densities, which indicates a variety of planetary compositions ranging from 
a water world to an iron core \citep[e.g.,][]{Howe14,Lopez14}. 
The observations of planetary atmospheres revealed the importance of the C/O ratio in estimating 
compositions and formation processes of planets \citep[e.g.,][]{Madhusudhan11}.
Furthermore, the advance in observations of protostellar discs started revealing the distribution of 
refractory and volatile molecules across the disc as well as the locations of condensation fronts of some volatiles 
\citep[e.g.,][]{Mathews13,Qi13}.
The future observations will place further constraints on compositions of protostellar discs and planets 
and help us understand the Solar System as one of the planetary systems.

The compositions of planets may be largely determined by the primordial compositions 
of a protostellar disc.
However, there are also good indications that dynamical evolution of giant planets could significantly 
affect the compositions of terrestrial planets.
Two major pathways for giant planets to influence the formation and survival of rocky planets 
are migration in the disc and dynamical instabilities among them. 
Migration of a giant planet through the inner disc could be potentially hazardous to 
formation of terrestrial planets \citep{Armitage03}. 
Some numerical studies showed that small bodies interior to a migrating giant planet could 
survive either by being shepherded toward the star or by being scattered outward 
\citep[e.g.,][]{Zhou05,Fogg05}, though the survival rate of such bodies depends on 
the migration rate of the massive body \citep[e.g.,][]{Ward95,Tanaka99,Edgar04}.
\cite{Raymond06} showed that such migration could mix the disc materials radially and 
create a variety of planetary systems with both dry and water-rich planets.

Dynamical instability of giant planets, on the other hand, could be a major threat 
during the final assembly of terrestrial planets, which is expected to occur 
after the disc's dispersal.
Indeed, the presence of dynamically active giant planets could drive building blocks 
of rocky planets to merge with the central star due to the sweeping of secular resonances, 
or to be ejected out of the system due to scattering by giant planets 
\citep[e.g.,][]{Veras05,Veras06,Raymond11,Raymond12,Matsumura13}.
\cite{Raymond12} showed that some outer disc materials 
could be scattered into the inner disc by giant planets, creating a water-rich planet. 
Thus, both migration and dynamical instability of giant planets could help moving the building blocks of 
terrestrial planets from their initial locations.

In this paper, we explore how dynamical evolution of giant planets affects the compositions of 
terrestrial planets by investigating the Solar System formation.
There are two major models for the formation of rocky planets in the Solar System --- a classical formation 
model and a so-called Grand-Tack model.
In the classical model, giant planets are assumed to have present masses and current orbital parameters. 
\cite{Chambers01} studied such a scenario and reproduced various features of the Solar System planets 
including the formation timescale (within $200\,$Myr), the number of planets (typically 3-4), the fraction of mass in the largest planet compared to the others, 
and wide orbital spacings compared to giant planets. 
However, this model disagreed with the observed systems in some ways --- eccentricities and inclinations 
were higher than the terrestrial planets in the Solar System, and masses of Mercury and Mars analogues were 
higher than those planets.
For the former point, the dynamical friction due to a large number of small planetesimals 
was proposed to be a potential solution to high eccentricities and inclinations of simulated rocky planets 
\citep[e.g.,][]{Chambers01,OBrien06,Raymond06}.
For the latter point, \cite{Hansen09} showed that the mass distribution of the Solar System could be reproduced 
when the disc materials were concentrated in a very narrow region near Venus and Earth orbits.
However, this work did not explain how such a high density region could be created. 

\cite{Walsh11} proposed a so-called Grand Tack model, which could at least partially explain the creation 
of a high density region in the inner disc.
In the first step of their model, Jupiter is assumed to form before Saturn, clear the gas in an annulus whose width is comparable to its Hill radius, 
and undergo inward Type II migration \citep[e.g.,][]{Ward97}. 
As Jupiter migrates, it scatters some materials out of the system and shepherds some others down to the inner disc, creating 
an enhanced density region for terrestrial planet formation. 
Once Saturn grows to a critical mass, it is assumed to open a partial gap, migrate rapidly, catch up with Jupiter, 
and get trapped in a mean-motion resonance. 
In this process, these two giant planets share a gap in the disc together, 
which effectively means that Jupiter receives an angular momentum from the inner disc 
and Saturn loses an angular momentum to the outer disc while they exchange an angular momentum with each other.   
Since Jupiter is more massive than Saturn, the pair gain angular momentum overall and migrate outward \citep[e.g.,][]{Masset01,Morbidelli07,Pierens08,Pierens11}. 
In other words, the planets reverse their migration and they 'tack'. 
Once giant planets move away and the gas disc dissipates, 
terrestrial planets could form via collisions between embryos and planetesimals, 
free from close encounters with giant planets. 
Their work successfully reproduced the mass distribution of rocky planets in the Solar System including Mars analogues 
and also explained the compositional differences across the asteroid belt.
However, it is unclear whether each step of this scenario would work correctly. 
For example, the particular configuration and outward migration favoured by Grand Tack is only supported for a specific set of initial conditions 
\citep{DAngelo12} and is not universal \citep{Zhang10}.
There may also be other pathways to produce such a high density region \citep[e.g.,][]{Izidoro14}.
Furthermore, the Grand Tack model needs to explain how the innermost disc ($<0.5\,$AU) is cleaned up.

Here, we will numerically study both classical and Grand Tack formation models and explore the differences 
in elemental abundances of planets. 
In Section~2, we describe chemical models as well as numerical methods and the initial conditions. 
In Section~3, we present results for different formation models and compare their results with each other.
In Section~4, we discuss and summarise our work.

Throughout this paper, we use the following terminologies for various elements.
The standard classification scheme identifies the major rock-forming elements (e.g., $\rm Mg$, $\rm Fe$, $\rm Si$, and $\rm Ni$), 
and categorises the other elements into four groups according to their volatility \citep[e.g.,][]{McDonough03,Davis06}.
The refractory elements (e.g., $\rm Ca$, $\rm Al$, $\rm Ti$ in Table~\ref{tab2}) are the ones 
with equilibrium condensation temperatures higher than the major elements, 
and the critical temperature is $>1335\,$K for gas of solar composition at a pressure of $10^{-4}\,$atm. 
The moderately volatile elements (e.g., $\rm Cr$, $\rm Li$, $\rm Na$, $\rm K$, and $\rm P$) 
are the ones with $1335-665\,$K for the same condition, 
and the highly volatile elements (e.g., $\rm Cd$, and $\rm Pb$, not included in Table~\ref{tab2}) are the ones with $<665\,$K.
The atmophile elements (e.g., $\rm H$, $\rm C$, $\rm N$, and $\rm O$ in Table~\ref{tab2} as well as noble gases) condense 
at and below the condensation temperature of water ($180\,$K at $10^{-4}\,$atm).
%
%
\section{Numerical methods and initial conditions}\label{methods}

\subsection{Dynamical Model}\label{dyn}
All the simulations in this paper start with small planetesimals, nearly Mars-mass embryos as well as 
Jupiter- and Saturn-like planets.
We do not include Uranus and Neptune in any of the simulations, 
because they do not have as direct and significant effects as Jupiter and Saturn on terrestrial planet formation.
%

In all Grand Tack simulations, we do not take account of the effect of Saturn's mass growth either, because 
\cite{Walsh11} showed that the overall results were largely determined by the tack location and the initial location of Jupiter, 
and that other effects including the presence of Uranus and Neptune, the radial evolution of Saturn, and the migration timescale 
did not change the results much.
We assume that Jupiter has its current mass and is initially placed on a near-circular orbit at $3.5\,$AU 
and that a fully-grown Saturn is placed in the 2:3 resonance with Jupiter at $4.5\,$AU.
Grand Tack models have been explored by assuming the tack location of $1.5\,$AU \citep[e.g.,][]{Walsh11,Jacobson14}. 
However, as pointed out by \cite{Walsh11}, the tack location does not need to be $1.5\,$AU and 
this should be constrained by observed properties of the Solar System planets.
Here, we test two sets of Grand Tack simulations with different tack locations --- 
one at $1.5\,$AU (hereafter GT1.5AU) and the other at $2.0\,$AU (hereafter GT2.0AU).

The migration of the gas giants was mimicked through fictitious forces as 
detailed in \cite{Walsh11}. 
In GT1.5AU, during the first $0.1\,$Myr, 
Jupiter and Saturn migrate from the initial locations ($3.5\,$AU and $4.5\,$AU) to the tack locations 
($1.5\,$AU and $2.0\,$AU), respectively. 
For the next $0.4\,$Myr, Jupiter and Saturn migrate out to $5.2\,$AU and $6.9\,$AU, and 
end up in appropriate initial conditions for the NICE model \citep{Tsiganis05,Morbidelli05,Gomes05}. 
While the gas disc is around, the planetary embryos experience tidal damping of 
their eccentricities and inclinations from the gas disc, and the planetesimals suffer that from gas drag. 
After $0.5\,$Myr, the disc is assumed to be dispersed completely and bodies just experience gravitational interactions 
with one another.
For gas drag purposes, we assume a planetesimal size is $50\,$km.
The migration in GT2.0AU is set up in a similar manner to GT1.5AU.

For the classical formation model, we follow the approaches by \cite{Chambers01} and \cite{OBrien06} 
and assume that Jupiter and Saturn have current masses, radii, and orbital properties. 
We will call these simulations Eccentric Jupiter and Saturn (hereafter EJS) runs, following \cite{OBrien06}. 

We employ another set of simulations, where Jupiter and Saturn are on circular and nearly aligned orbits,  
and call them Circular Jupiter and Saturn (hereafter CJS) runs.
Specifically, Jupiter and Saturn have semimajor axes of $a_J=5.45\,$AU and $a_S=8.18\,$AU, 
eccentricities of $e_J=e_S=0$, and inclinations of $i_J=0$ and $i_S=0.5^{\circ}$ initially.
These conditions mimic the Nice Model that leads to the late heavy bombardment 
in the time comparable to the observational estimate \citep[$\sim700\,$Myr,][]{Gomes05}. 
In both EJS and CJS simulations, Jupiter and Saturn do not migrate in the disc.

To compare the effects of different formation models on compositions of planets, 
we use a similar distribution of embryos and planetesimals for all the cases.
All simulations in this paper assume the equal-mass embryos, which agrees better with a pebble-accretion scenario \citep{Morbidelli15} 
rather than a more traditional oligarchic-growth scenario \citep{Kokubo98}.
We will discuss the differences between these two scenarios in the follow-up paper \citep{Brasser16pre}.

We also assume that the total mass ratio of embryos and planetesimals 
in the inner disc ($0.3-3\,$AU) is $1:1$. 
More specifically, the total inner disc mass is $3.74\,M_{\oplus}$ and 
half the mass is in 20 embryos with a mass comparable to Mars ($9.34\times10^{-2}\,M_{\oplus}$), 
and the other half is in 1571 planetesimals.
Such a bimodal distribution was shown to produce the number of planets similar to the Solar System ($3.5\pm0.5$) 
compared to more uniform mass distributions \citep[$2.9\pm0.6$,][]{Chambers01}. 
We do not explore different total mass ratios for embryos and planetesimals in this work.  
However, \cite{Jacobson14} studied such effects 
and reported that the larger mass ratios tend to lead to higher excitation and less concentration of planets 
compared to terrestrial planets in the Solar System.

All simulations in this paper also include the outer planetesimals unless it is noted otherwise. 
These consist of $500-650$ planetesimals with an individual mass of $1.2\times10^{-4}\,M_{\oplus}$, 
following \cite{Rubie15}.
The outer planetesimals are distributed over $5-9\,$AU for GT, $3-8\,$AU for CJS, and $3-9.5\,$AU for EJS models. 
The eccentricities and inclinations of both embryos and planetesimals are randomly chosen 
from a uniform distribution between $0-0.01$ and $0^{\circ}-0.5^{\circ}$ as in \cite{Chambers01} and \cite{OBrien06}.
The other angular orbital elements were randomly chosen from $0^{\circ}-360^{\circ}$.
%

The system consisting of gas giants, planetary embryos and planetesimals is integrated with the SyMBA software package 
\citep{Duncan98}, with a time step of $0.02\,$yr for $200\,$Myr. 
We compute the mutual gravity between gas giants and embryos, but the planetesimals are treated as test particles which 
are only perturbed by other bodies and do not affect others. 
The approximation is used to keep the CPU time within reasonable limits, 
and is justified because Jupiter destroys the disc beyond $1\,$AU in $0.1\,$Myr. 
Planets and planetesimals are removed once they are farther than $100\,$AU from the Sun (whether bound or unbound) 
or when they collide with a planet or venture closer than $0.1\,$AU from the Sun.  
We run 8 simulations without and 8 simulations with outer planetesimals for both GT1.5AU and GT2.0AU, 
while we run 6 simulations with outer planetesimals for both CJS and EJS.

\subsection{Chemistry Model}\label{chem}
Knowing the primordial compositions of a protostellar disc is a starting point of estimating the compositions of 
planets in the context of planet formation.
The primordial compositions of the protostellar disc in the Solar System can be inferred from the 
solar photosphere abundances, because a protostellar disc is formed around a protostar as 
a molecular core gravitationally collapses and thus should be formed from the same materials as the star.
Such a collapse and a following disc accretion can heat the disc up intensely and reset the 
chemistry occurred in the molecular cloud core.
In this picture, the elements condense as the disc cools, by following the condensation sequence. 
The assumption is supported by the elemental abundance pattern of meteorites and 
bulk terrestrial planets \citep[e.g.,][]{Davis06,Pontoppidan14}. 
The abundances of these bodies are depleted with respect to the most primitive bodies, CI chondrites, 
which have abundances comparable to the solar photosphere except for the most volatile elements \citep[e.g.,][]{Lodders03}, 
and the abundances overall decrease with the $50\%$ condensation temperatures of these elements \citep[e.g.,][]{Cassen96,Davis06}.
The equilibrium condensation in an evolving protostellar disc has been shown to explain 
such a depletion pattern well \citep{Cassen96,Cassen01}. 
The assumption also predicts that refractory minerals such as Ca-, Al-, and Ti-oxides would be the first ones to 
condense, which agrees with the observation that Calcium-Aluminum Inclusions (CAIs) are the oldest objects 
in the Solar System \citep[e.g.,][]{Ciesla06,Pontoppidan14}.
%

However, in particular in the outer region of a disc, the temperature is expected to 
never reach the sublimation temperatures of some of the most volatile species \citep[e.g.,][]{Messenger06}.
Thus, in such regions, presolar grains from a molecular cloud core could survive. 
The assumption is supported by the widespread presence of presolar grains \citep[e.g.,][]{Davis06,Pontoppidan14}.  
The picture of the primordial compositions of a proto-solar disc is further complicated 
because the path to equilibrium condensation can be altered by various processes 
such as transport of chemical species and transient heating events \citep{Ciesla06,Ebel06}.
Overall, most regions of the disc show the evidence of both the inheritance from the cloud core and the disc processing 
after resetting the previous chemical features, while the latter plays a more important role in the inner disc \citep[$<20\,$AU,][]{Pontoppidan14}.
In this paper, we are interested in the disc compositions and thus the compositions of planet building blocks 
within $10\,$AU as shown in Section~\ref{dyn}.  
Therefore, we will ignore the effects of presolar grains and other disc processing effects such as shocks and lightning, 
and focus on the equilibrium condensation model.

%
\cite{Bond10a} followed the idea by \cite{Cassen96} and applied the chemical equilibrium model to planet formation 
simulations by \cite{OBrien06}. 
For given temperature and pressure in a certain region of a disc, they determined 
which gaseous species condense out of the disc by using a commercial software HSC chemistry, 
and used them as compositions of planetesimals and embryos in that region.  
They found that the abundances of refractory and moderately volatile elements in planets 
agree well with the estimated values of terrestrial planets in the Solar System.   
A notable disagreement between their model and the actual planets is a lack of accretion of atmophile elements C and N.   
The abundances of these elements in the Earth are very small (see Table~\ref{tab6}), 
but they are two of six major biogenic elements (H, C, N, O, S, and P) and thus are crucial for the origin of life.  
In \cite{Bond10a}, only C and N out of these did not accrete onto planets, 
because they have low condensation temperatures 
and major species such as CO and ${\rm NH_3}$ were not included as solid species in their calculations.
\cite{Marboeuf14a} developed a model that calculates the compositions of ices in a protostellar disc.  
We will take account of the effects of icy species including C and N by following their work as described below.
%
%

Another assumption made in \cite{Bond10a} is that the compositions of building blocks of planets 
(e.g., planetesimals and planetary embryos) are determined at a single instant in the disc's evolution.
\cite{Moriarty14} recently improved this assumption and took account of the sequential condensation 
effect by converting a fraction of a disc into planetesimals, evolving the disc, and recalculating 
the equilibrium chemical composition at every timestep.   For the solar system $\rm C/O$ value ($0.54$), 
they found that the planetary compositions are similar for both equilibrium and sequential condensation assumptions.  
On the other hand, for C-rich systems with $\rm C/O > 0.8$, they showed that the C-rich planetesimals and planets form 
over a wider orbital radii.  
Since we focus on the Solar System in this study, 
we will determine the initial compositions of building blocks at a single instant, 
as in \cite{Bond10a} and \cite{Elser12}. 
This will also help us clearly see a difference between the iceline evolution and the dynamical effects.

\begin{deluxetable}{lc}
\tablecaption{Elemental Abundances used for HSC chemistry \label{tab1}}
\tablecolumns{2} 
\tablehead{ 
\colhead{Element Name} &  
\colhead{Abundance (mol)}}
\startdata
H  & $1.00\times10^{12}$ \\
He & $8.51\times10^{10}$ \\
C  & $2.45\times10^{8}$ \\
N  & $6.03\times10^{7}$ \\
O  & $4.57\times10^{8}$ \\
Na & $1.48\times10^{6}$ \\
Mg & $3.39\times10^{7}$ \\
Al & $2.34\times10^{6}$ \\
Si & $3.24\times10^{7}$ \\
P  & $2.29\times10^{5}$ \\
S  & $1.38\times10^{7}$ \\
Ca & $2.04\times10^{6}$ \\
Ti & $7.94\times10^{4}$ \\
Cr & $4.37\times10^{5}$ \\
Fe & $2.82\times10^{7}$ \\
Ni & $1.70\times10^{6}$
\enddata
\tablecomments{Column 1: name of the element, and 
column 2: elemental abundance in moles.
These values are scaled with respect to H to use in HSC chemistry. 
All values are taken from \cite{Bond10a}.}
\end{deluxetable}
For refractory and moderately volatile elements, we follow previous studies and use 
the HSC Chemistry that determines the equilibrium composition 
by iteratively minimizing the system's Gibbs free energy 
(e.g., an isothermal-isobaric system in equilibrium --- $\Delta T=0$ and $\Delta P =0$).
The underlying assumption here is that the disc evolves slower than the time required to 
reach the chemical equilibrium so that the temperature and the pressure are approximately constant. 
This assumption is likely valid except for the icy midplane beyond the iceline, 
because the chemical reaction timescale is expected to be short \citep[$\sim10^4\,$yr or shorter][]{Woitke09} 
compared to the disc's evolution timescale.

\begin{deluxetable}{lc}
\tabletypesize{\footnotesize}
\tablecaption{Condensation Temperatures for Solid Species included in HSC calculations \label{tab2}}
\tablecolumns{2} 
\tablehead{
\colhead{Solid Name} &  
\colhead{Temperature (K)}}
\startdata
$\rm CaAl_{12}O_{19}$ & $1790.5$ \\
$\rm CaS$ & $1633.5$ \\ 
$\rm CaTiO_3$ & $1633.5$ \\
$\rm Ca_2Al_2SiO_7$ & $1452.9$ \\
$\rm CaMgSi_2O_6$ & $1343.3$ \\
$\rm Fe$ & $1343.3$ \\
$\rm MgSiO_3$ & $1319.3$ \\
$\rm Mg_2SiO_4$ & $1319.3$ \\
$\rm Ni$ & $1295.7$ \\
$\rm Fe_3P$ & $1191.1$ \\
$\rm Cr_2FeO_4 $ & $1162.8$ \\
$\rm MgAl_2O_4$ & $1135.1$ \\
$\rm Ti_2O_3$ & $1135.1$ \\
$\rm Ca_3(PO_4)_2$ & $876.4$ \\
$\rm NaAlSi_3O_8$ & $850.5$ \\
$\rm FeS$ & $723.0$ \\
$\rm FeSiO_3$ & $723.0$ \\
$\rm Fe_2SiO_4$ & $410.7$ \\
$\rm Fe_3O_4$ & $305.9$ \\
$\rm Al_2O_3$ & $227.8$ \\
$\rm Mg_3Si_2O_5(OH)_4$ & $227.8$ \\
$\rm H_2O$ & $175.9$ 
\enddata
\tablecomments{Column 1: names of the solid species, and 
column 2: condensation temperatures (K) for $t_{\rm init, \, disc}=0.2\,$Myr.}
\end{deluxetable}

As in the previous studies, we start with the solar photosphere abundances of 16 elements in Table~\ref{tab1}, 
and calculate their equilibrium condensation at each radius of a disc model. 
This assumption is reasonable because the Sun and planets are formed out of the same molecular cloud core, and because 
the stellar pollution effect is expected to be limited.
The initial compositions of planetesimals and embryos are determined for temperatures and pressures corresponding to 
a Chambers' disc model \citep{Chambers09}.
We use 12 different initial disc ages --- $t_{\rm init, \, disc}=0$, $0.05$, $0.1$, 
$0.2$, $0.3$, $0.4$, $0.5$, $0.6$, $0.7$, $0.8$, $0.9$, $1\,$Myr.
Note that $t_{\rm init, \, disc}$ does not indicate the time evolution of a disc, and simply determines the initial 
temperature and pressure profiles of the disc. 
We assume that all planetesimals and embryos form at $t_{\rm init, \, disc}$ and determine their initial compositions at this instant.
Table~\ref{tab2} summarises the condensation temperatures for solid species that condense for $t_{\rm init, \, disc}=0.2\,$Myr. 

For atmophile elements, we follow \cite{Marboeuf14a} and take account of both condensation and clathrate hydrate formation on the surface of refractory grains \citep[also see][]{Marboeuf14b,Thiabaud14}.
A clathrate hydrate is a ``cage'' compound where a gas molecule 
such as $\rm H_2S$, $\rm CH_4$, $\rm CO$, and $\rm N_2$ is trapped inside the water ice lattice.
The clathrate formation could occur at higher temperatures than pure condensation 
and thus may help incorporating ices into building blocks of planets.
They assume solar abundances, and consider condensation and clathrate formation of 
most abundant atmophile species in ISM and comets of the solar system: 
$\rm H_2O$, $\rm CO$, $\rm CO_2$, $\rm CH_3OH$, $\rm CH_4$, $\rm N_2$, $\rm NH_3$, and $\rm H_2S$.
In their model, these species either condense or form clathrates when the equilibrium pressures 
become lower than or equal to their partial pressures (i.e., total pressure of the disc times the abundance of the species).
They have also taken account of the effects of refractory organics.
However, they found that the ice/rock ratio of observed comets is more consistent with 
that of icy planetesimals estimated without these organics.  
Thus, we will focus on the cases without refractory organics here.  

\begin{figure*}
\plotone{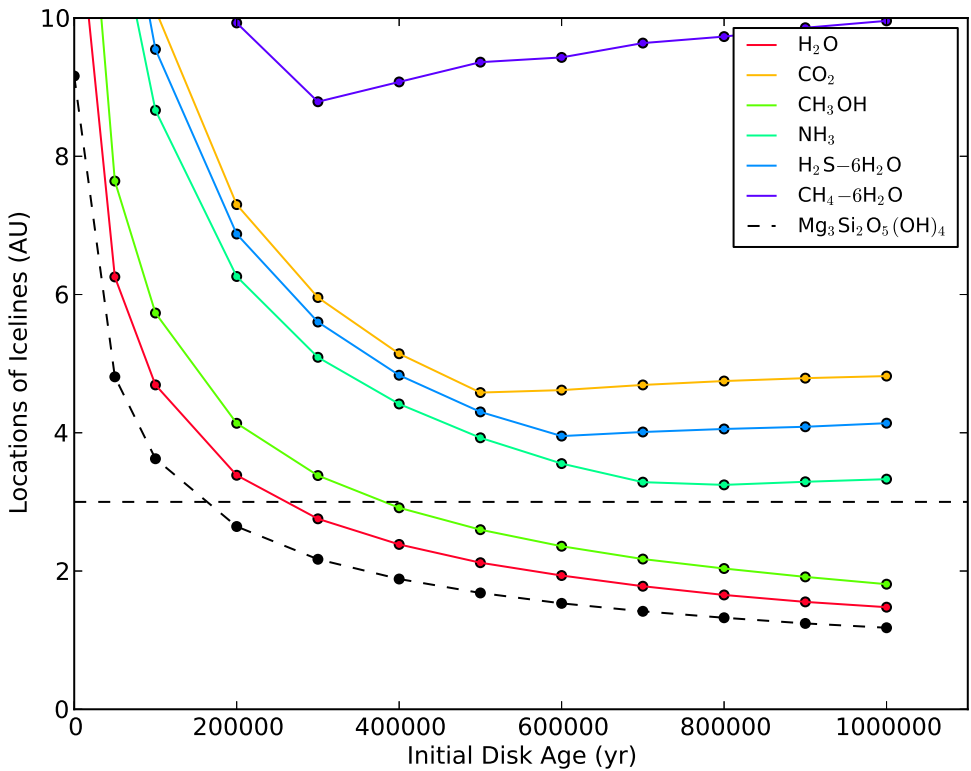}
\caption{Evolution of icelines of some highly volatile molecules in a Chambers' disc model.  
The dashed line is drawn at $3\,$AU, where inner and outer planetesimals are separated in our default setup. 
\label{icelines}}
\end{figure*}
Figure~\ref{icelines} shows the location changes of icelines for these molecules in a Chambers' disc model. 
The iceline for serpentine ($\rm Mg_3Si_2O_5(OH)_4$) is also plotted for comparison. 
The closest iceline is for water, but icelines for carbon and nitrogen such as $\rm CH_3OH$ and $\rm NH_3$ 
also move near the rocky planet formation region as the disc evolves and the temperature decreases.
The icelines for $\rm CO$ and $\rm N_2$ and their clathrates are beyond 10 AU all the time.
The figure indicates that, as the disc evolves, the building blocks of terrestrial planets are more likely to include 
atmophile species such as $\rm C$ and $\rm N$.
Note that, in a Chambers' disc model, the outer disc is dominated by stellar irradiation rather than by viscous heating, 
and thus the outer disc temperature stays the same for constant stellar properties as the disc pressure decreases.  
The irradiation dominated region also expands into the inner region as the disc evolves \citep[see Figure~1 of][]{Chambers09}.
This is the reason why some of the icelines (e.g., $\rm CH_4$ clathrates) move inward and then outward with time.
Also note that the standard planet formation expects a factor of a few increase in the surface mass density 
beyond the water iceline, and thus the mass of embryos \citep[e.g.,][]{Kokubo00,Morbidelli15}.
However, we do not change the mass of embryos across the water iceline for the ease of the comparison of models,  
so that all the embryos have the same mass in all the disc models we test.

%
\section{Results}\label{results}
\subsection{Dynamical Results}
Before presenting the elemental abundances of planets generated in our simulations, 
we first discuss the dynamical outcomes of Grand Tack, CJS, and EJS models. 
These models have been explored by other authors previously \citep[e.g.,][]{Walsh11,Jacobson14,OBrien06,OBrien14}, 
and our results overall agree with theirs. 
However, we discuss the outcomes of these models and compare them with each other, 
because they will give us clues to understand the compositional differences of planets 
that we discuss in Section~\ref{abundance_planets}.
 
\subsubsection{Overall Results} 
\begin{figure*}
\plotone{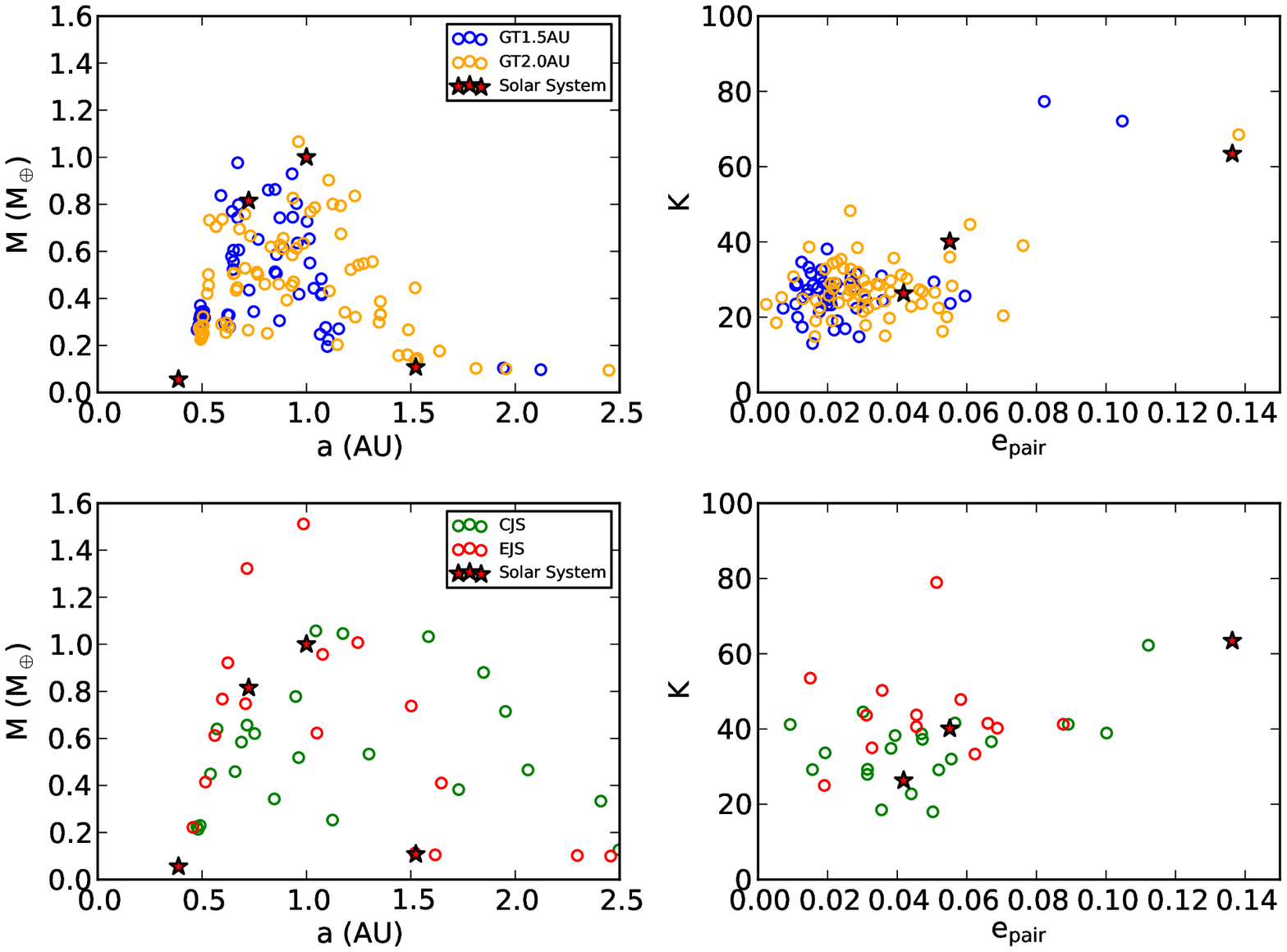}
\caption{The comparison of results of GT1.5AU and GT2.0AU (top panels) and CJS and EJS (bottom panels). 
The left panels show the mass distributions of planets and right panels show the 
separation of the neighbouring pairs in the mutual Hill radius as a function of the average 
eccentricities of the pairs.
The red stars represent corresponding results for Mercury, Venus, Earth, and Mars.  
\label{scatter_aeim1}}
\end{figure*}

All of the models generate a few planets per system in $\sim10-100\,$Myr, 
which is consistent with the Earth's formation timescale estimated from radioactive elements \citep[$10-150\,$Myr, e.g.,][]{Halliday06,Kleine09}.
The overall agreement with properties of the Solar System is shown in Figure~\ref{scatter_aeim1}, 
and the average properties are summarised in Table~\ref{tab3}.

%
%
In all of these models, the mass distributions have a peak near Venus and Earth locations. 
However, for the CJS model, the mass distribution is much broader than the other models, 
which indicates that the circular giant planet assumption is inconsistent with the 
terrestrial planet formation in the Solar System. 
The mass distributions from GT1.5AU and GT2.0AU are broadly similar to each other. 
For the further tack location ($2.0\,$AU), we find that 
the mass distribution becomes slightly broader than the default case 
and that the location of the peak also shifts slightly outward.
The production of Mars analogues is more common for GT2.0AU.  
We will investigate the effects of tack locations on the planetary mass distribution 
in our future paper \citep{Brasser16pre}. 

A general trend for the eccentricity and inclination distributions is similar for all the models as well.
For GT1.5AU and GT2.0AU, the average eccentricities are $0.024$ and $0.033$ while the average inclinations are 
$1.3^{\circ}$ and $1.5^{\circ}$, respectively. 
These are lower than the values for terrestrial planets in the Solar System ($e_{\rm ave}\sim0.096$ and $i_{\rm ave}=5.01^{\circ}$), and thus 
these planets have more circular and aligned orbits. 
The eccentricities and inclinations of the terrestrial planets could increase as 
giant planets go through the NICE-model phase of resonance crossing and migration \citep{Brasser13}.
Both CJS and EJS models also have lower average eccentricities than the Solar System, 
and lower or comparable average inclinations (see Table~\ref{tab3}).

All the models also reproduce the overall trend of the separation of the neighbouring pairs as a function of the 
average eccentricities of the pairs, where the separation increases with the average eccentricity.   
To calculate the separation $K$ in a mutual Hill radius, 
we define the mutual Hill radius as the average of the Hill radii of two planets. 
For the Solar System, the separation of the neighbouring pairs is 
$K=63.4$, $26.3$, and $40.1$ from inward to outward, and the average value is $K_{\rm ave}=43.3$ (see Table~\ref{tab3}).
For GT models, the final planetary systems tend to be more compact than the Solar System, 
and the average separation is $K_{\rm ave}\sim28$ for both GT1.5AU and GT2.0AU.
The average separation for CJS is larger than GT ($K_{\rm ave}\sim34$), and 
that for EJS ($K_{\rm ave}\sim45$) is comparable to the Solar System value.
The larger average separations for CJS and EJS models compared to GT models 
are consistent with their slightly higher eccentricities and inclinations.
%

\subsubsection{Dynamical fates and the mixing of the inner disc}\label{fates_scattering}
\begin{figure*}
\plottwo{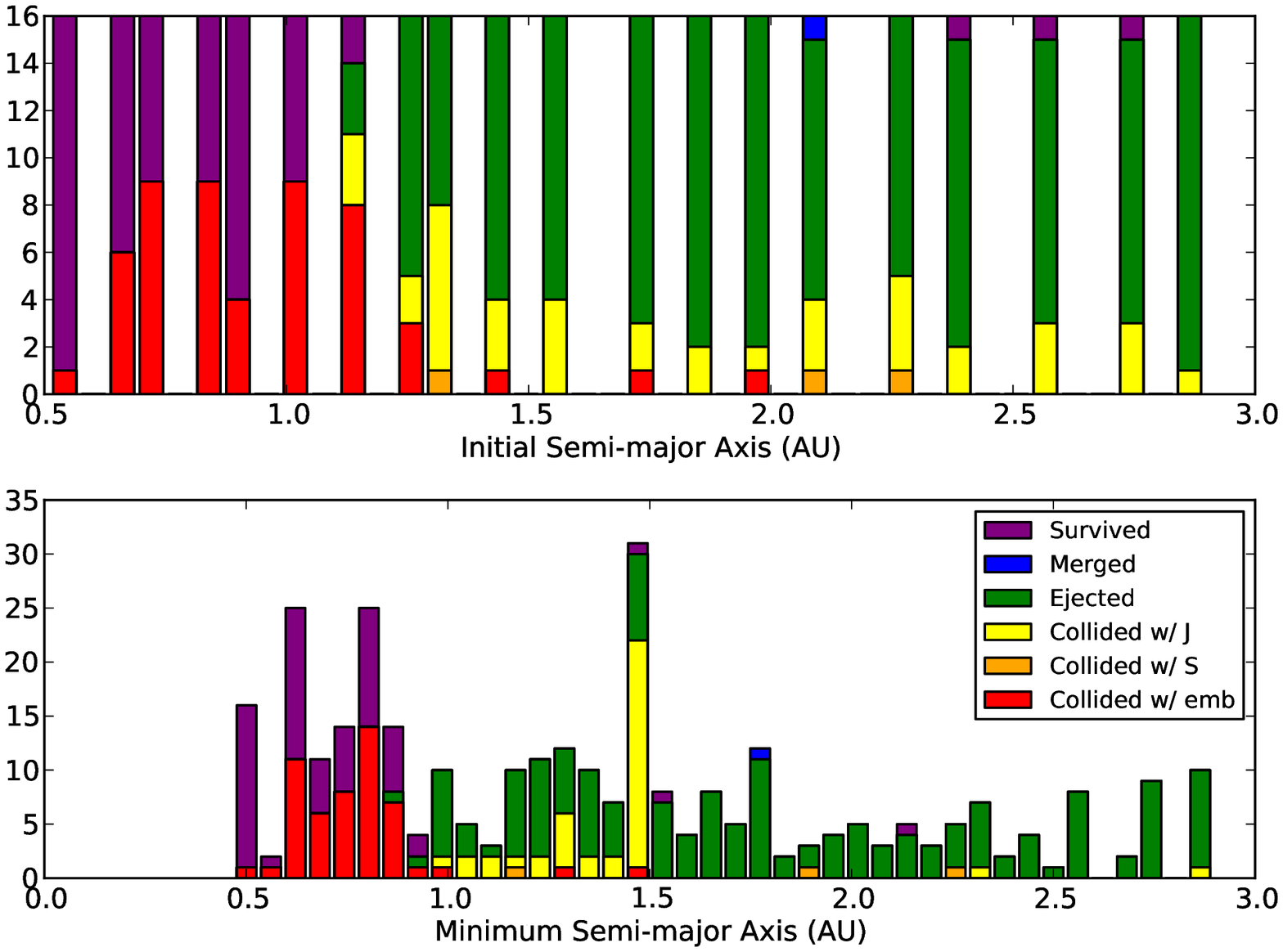}{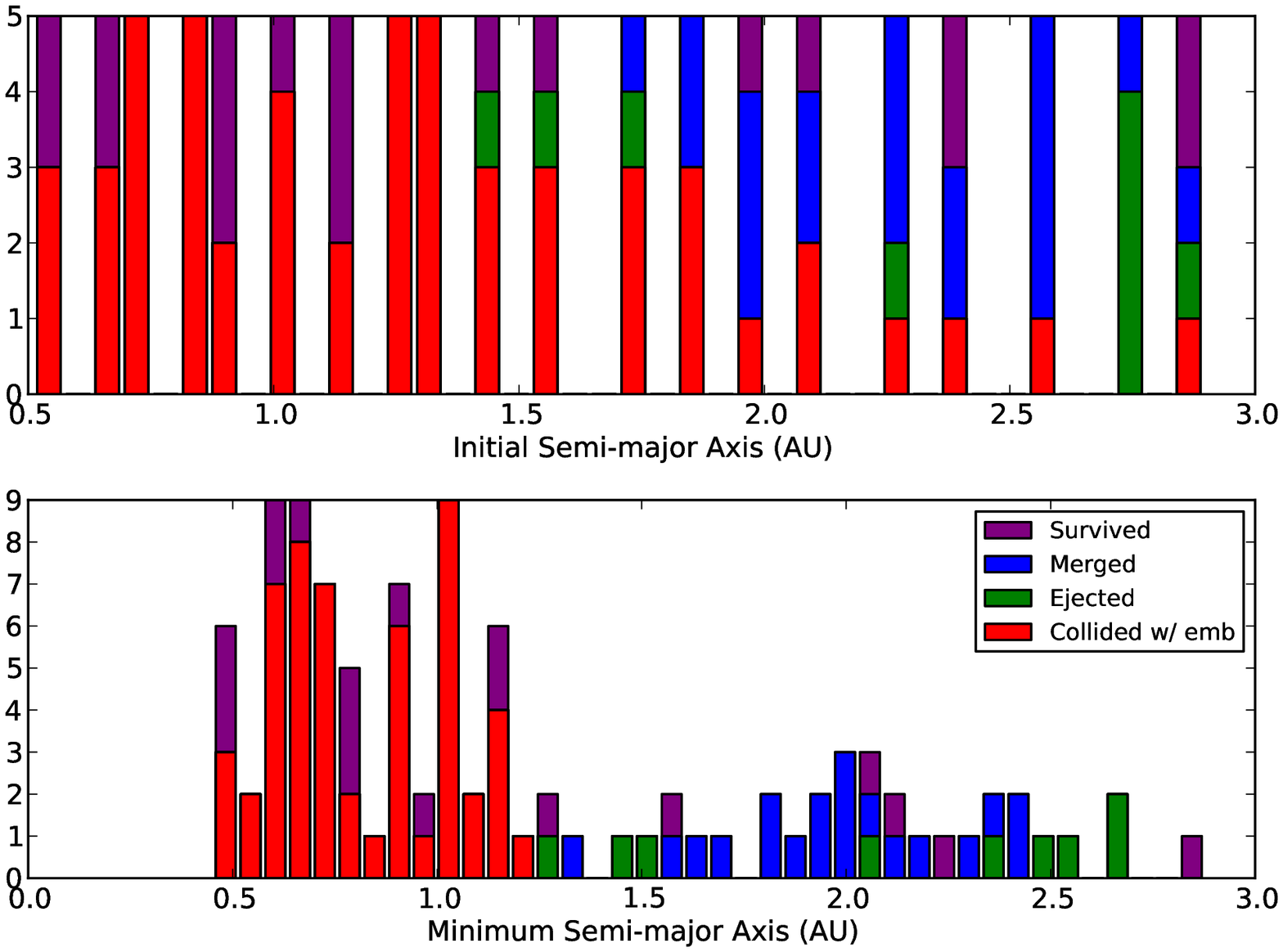}
\caption{The comparison of dynamical outcomes of embryos of GT1.5AU (left panels) and EJS models (right panels). 
The top panels show these outcomes as a function of the initial semimajor axis of an embryo, 
while the bottom panels are plotted against the minimum semimajor axis reached during the simulation.
\label{fates_emb}}
\end{figure*}

Although the distributions of planetary masses are similar for GT and EJS models (see Figure~\ref{scatter_aeim1}), 
the dynamical fates of embryos are markedly different for these models as seen in Figure~\ref{fates_emb}. 
The breakdown of various fates of embryos are summarised in Table~\ref{tab3}.
In GT models, most survived embryos (purple) were initially within the tacking radius of Jupiter 
($1.5\,$AU or $2.0\,$AU), while in EJS and CJS models, survived planets come from a wider range of orbital radii. 
A difference in the initial locations of the survived embryos 
indicates a potential importance of tacking locations in determining their compositions. 
In Section~\ref{abundance_planets}, we will show that these differences do not lead to 
significant changes in elemental compositions of planets.
In all of these models, embryos colliding with other embryos (red) come from the similar regions to survived ones.
\begin{deluxetable*}{lccccccccc}


\tablecaption{Outcomes of runs for embryos. \label{tab3}}
\tablecolumns{10} 
\tablehead{ 
\colhead{Run Name} &  
\colhead{$e_{\rm ave}$} & \colhead{$i_{\rm ave}\,$(deg)} & \colhead{$K_{\rm ave}\,(R_{\rm mut,\,Hill})$} & 
\colhead{S} & \colhead{M} & \colhead{E} & \colhead{CE} &  
\colhead{CJ} & \colhead{CS}}
\startdata
GT1.5AU & $0.024$ & $1.26$ & $28.0$ & $19.7$ & $0.31$ & $50.3$ & $16.3$ & $12.5$ & $0.94$ \\
GT2.0AU & $0.033$ & $1.60$ & $28.1$ & $24.7$ & $0$ & $40.0$ & $29.4$  & $5.3$ & $0.63$ \\
CJS     & $0.045$ & $2.65$ & $34.8$ & $28.8$ & $0$ & $2.5$ & $68.8$  & $0$ & $0$ \\
EJS     & $0.042$ & $6.39$ & $45.0$ & $19.0$ & $19.0$ & $9.0$ & $53.0$  & $0$ & $0$ \\
Solar System & $0.096$ & $5.01$ & $43.3$ & $-$ & $-$ & $-$ & $-$ & $-$ & $-$    
\enddata
\tablecomments{Column 1: name of a set of runs, 
column 2-4: The average eccentricity, inclination, 
and separation of a neighbouring pair in the mutual Hill radius, respectively, 
column 5-10: Percentages of fates of embryos --- survived, merged with the central star, ejected out of the system, 
collided with another embryo, collided with Jupiter, and collided with Saturn, respectively. 
The information for the Solar System is added for comparison.}
\end{deluxetable*}

A characteristic difference between GT and EJS models is seen among the fates of 
the other embryos that are removed from the systems.
In GT simulations, some embryos collide with Jupiter and a few collide with Saturn, but the vast majority of them 
are ejected out of the system (see Table~\ref{tab3}).
The peak in a number of embryos collided with Jupiter is seen near the minimum semimajor axis comparable to 
the Jupiter's tack location, which indicates that most collisions occur near this radius.
Both ejection and collision in this scenario can be explained as a result of close encounters of embryos 
with the giant planets.
On the other hand, in EJS simulations, most other embryos merge with the central star, and only a few 
are ejected out of the system.  
The merger with the central star in EJS is likely due to secular resonances induced by Jupiter and Saturn 
which gradually increase eccentricities of embryos by keeping their semimajor axes constant \citep[e.g.,][]{Matsumura13}. 
The lack of collisions with Jupiter and Saturn for EJS makes sense because embryos in this model (and CJS) are far away from them 
and because scattering is more effective further from the star.

The corresponding result for CJS model is similar to that of EJS, and survived embryos 
come from all over the inner disc.
An interesting difference between EJS and CJS models is that most embryos in the CJS model 
collided with each other.  
There are no embryos that collide with Jupiter or Saturn, or merge with the central star, 
which indicates that planet formation proceeds without significant gravitational perturbations 
from the giant planets. 

\begin{deluxetable*}{lcccccccccccc}


\tablecaption{Outcomes of runs for planetesimals. \label{tab4}}
\tablecolumns{13} 
\tablehead{ 
\colhead{Run Name} &  
\colhead{$\rm S_{in}$} & \colhead{$\rm M_{in}$} & \colhead{$\rm E_{in}$} & 
\colhead{$\rm CE_{in}$} &  \colhead{$\rm CJ_{in}$} & \colhead{$\rm CS_{in}$} &
\colhead{$\rm S_{out}$} & \colhead{$\rm M_{out}$} & \colhead{$\rm E_{out}$} & 
\colhead{$\rm CE_{out}$} &  \colhead{$\rm CJ_{out}$} & \colhead{$\rm CS_{out}$}}
%
\startdata
GT1.5AU & $1.5$ & $1.9$ & $25.9$ & $65.5$ & $4.8$ & $0.4$  & $1.6$ & $0.2$ & $89.9$ & $0.9$ & $4.0$ & $3.3$ \\ 
GT2.0AU & $2.0$ & $4.6$ & $25.9$ & $61.8$ & $3.5$ & $0.3$  & $1.0$ & $0.8$ & $91.8$ & $0.8$ & $3.3$ & $2.3$ \\ 
CJS     & $3.0$ & $3.3$ & $33.3$ & $59.3$ & $0.9$ & $0.08$   & $5.1$ & $0.6$ & $85.2$ & $1.4$ & $6.9$ & $0.8$ \\ 
EJS     & $2.2$ & $41.3$ & $15.9$ & $40.2$ & $0.3$ & $0.04$   & $2.0$ & $1.5$ & $93.2$ & $0.05$ & $2.8$ & $0.4$  
\enddata
\tablecomments{Column 1: name of a set of runs, 
column 2-7: mass percentages of inner planetesimals that survived, merged with the central star, ejected out of the system, 
collided with another embryo, collided with Jupiter, and collided with Saturn, respectively,
column 8-13: the corresponding mass percentages of outer planetesimals.
}
\end{deluxetable*}

\begin{deluxetable*}{llcccc}
\tablecaption{Initial and final locations of planetesimals. \label{tab5}}
\tablecolumns{16} 
\tablehead{ 
\colhead{Run Name} & \colhead{$a_{\rm in}\,$(AU)} &  
\colhead{$a_{\rm fin}<1\,$(AU)} & \colhead{$a_{\rm fin}=1-2\,$(AU)} & \colhead{$a_{\rm fin}=2-3\,$(AU)} & \colhead{$a_{\rm fin}\geq3\,$(AU)} }
\startdata
GT1.5AU & $<1$ & $77.7$ & $14.3$ & $6.0$ & $2.0$ \\
 & $1-2$ & $58.3$ & $20.2$ & $7.0$ & $14.5$ \\
 & $2-3$ & $23.5$ & $7.3$ & $6.8$ & $62.4$ \\
 & $\geq3$ & $0.7$ & $0.4$ & $2.9$ & $96.0$ \\ 
GT2.0AU & $<1$ & $76.1$ & $15.3$ & $2.4$ & $6.3$ \\
 & $1-2$ & $48.1$ & $27.5$ & $4.8$ & $19.6$ \\
 & $2-3$ & $21.7$ & $11.8$ & $4.2$ & $62.2$ \\
 & $\geq3$ & $0.7$ & $0.5$ & $1.1$ & $97.8$ \\ 
CJS     & $<1$ & $70.5$ & $18.3$ & $2.1$ & $9.1$ \\
 & $1-2$ & $26.3$ & $35.3$ & $7.4$ & $31.1$ \\
 & $2-3$ & $9.5$ & $18.3$ & $10.4$ & $61.8$ \\
 & $\geq3$ & $0.5$ & $0.9$ & $1.2$ & $97.4$ \\ 
EJS     & $<1$ & $73.9$ & $16.4$ & $6.2$ & $3.5$ \\
 & $1-2$ & $26.2$ & $32.4$ & $31.6$ & $9.8$ \\
 & $2-3$ & $2.1$ & $9.5$ & $51.1$ & $37.2$ \\
 & $\geq3$ & $0$ & $0.03$ & $0.87$ & $99.1$ 
\enddata
\tablecomments{Column 1: name of a set of runs, 
column 3-6: mass percentages of planetesimals with the initial semimajor axis $a_{\rm in}<1\,$AU, $1-2\,$AU, $2-3\,$AU, and $\geq3\,$AU 
that end up with the final semimajor axis $a_{\rm fin}<1\,$AU, $1-2\,$AU, $2-3\,$AU, and $\geq3\,$AU. }
\end{deluxetable*}

\begin{figure*}
\plottwo{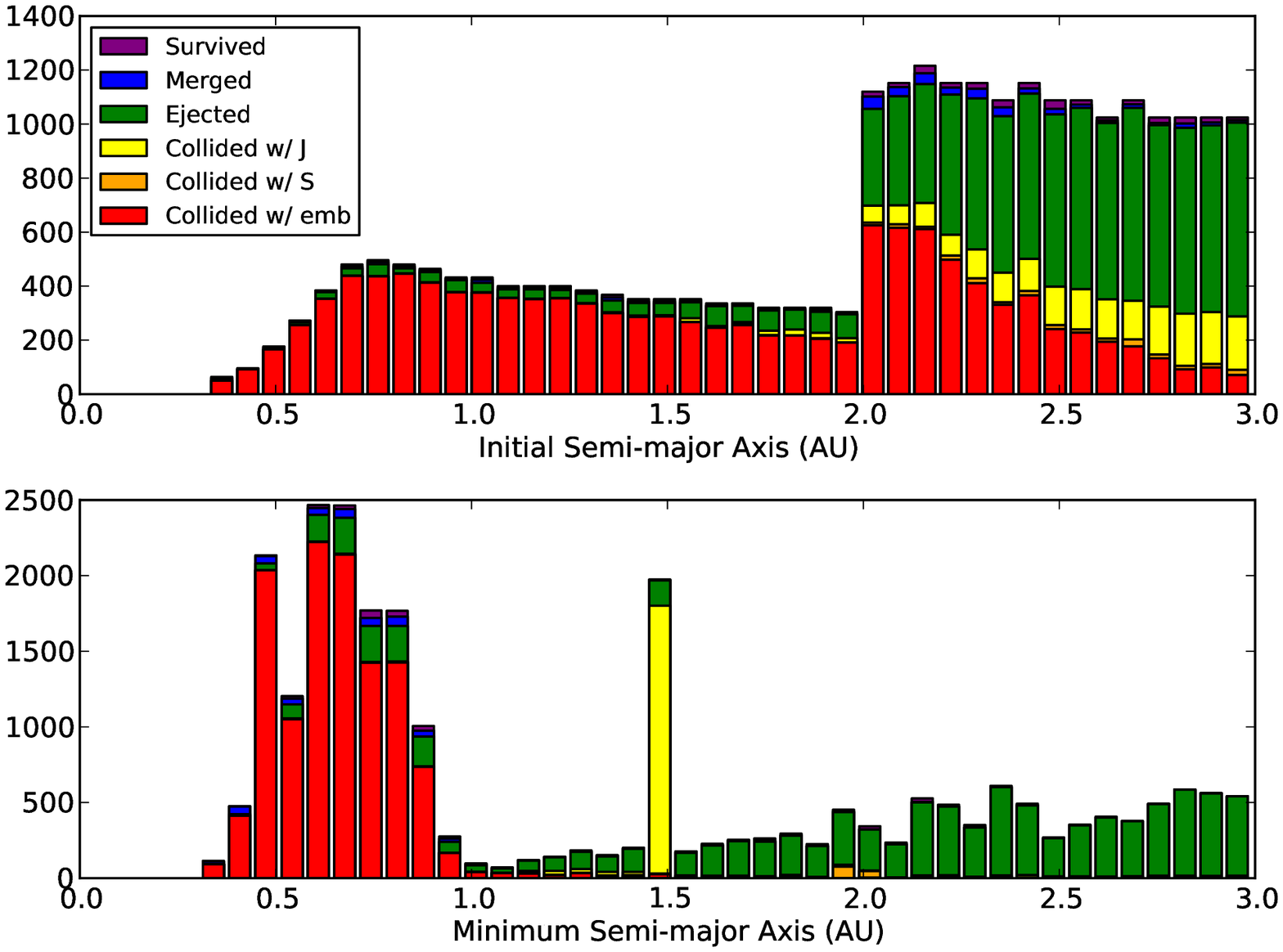}{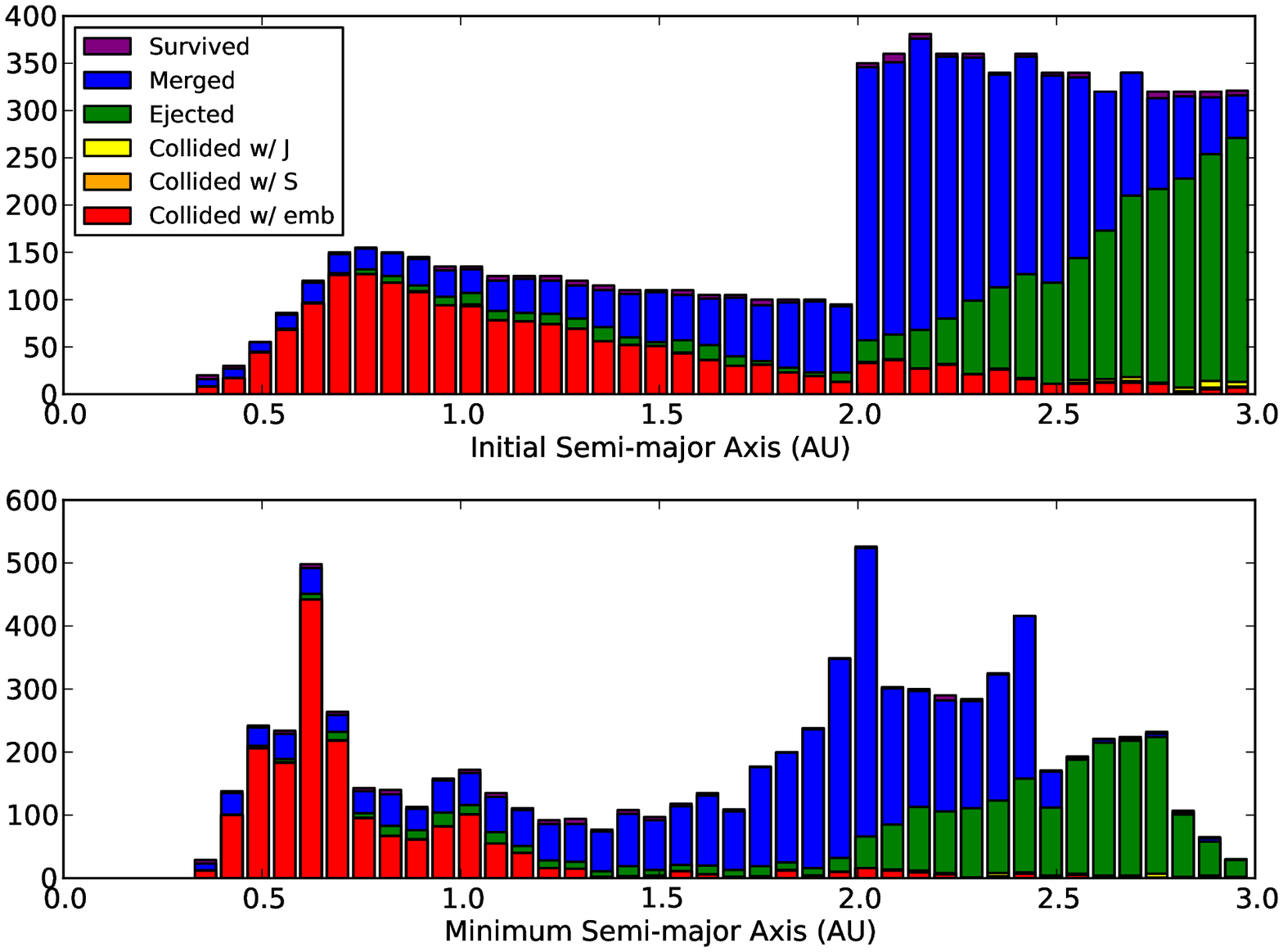}
\caption{The comparison of dynamical outcomes of inner planetesimals of GT1.5AU simulations 
(left panels) and EJS simulations (right panels). 
The top panels show these outcomes as a function of the initial semimajor axis of a planetesimal, 
while the bottom panels are plotted against the minimum semimajor axis reached during the simulation.
The jump in a number of planetesimals beyond $2.0\,$AU occurs because the planetesimal mass is a factor of 4 smaller 
there compared to within this radius.
\label{fates_plsml}}
\end{figure*}
The dynamical fates of inner planetesimals for GT1.5AU and EJS models are compared in Figure~\ref{fates_plsml}, 
and the breakdowns for all the models are summarised in Tables~\ref{tab4} and \ref{tab5}. 
Similar to embryos, merger with the central star ($\sim41\%$) is a more common outcome than ejection ($\sim16\%$)
for inner planetesimals in EJS model, but ejection is more common in GT models ($\sim26\%$ compared to a few percent). 
A notable difference from the trend of embryos is that planetesimals that collide with embryos are from all over the inner disc 
for GT while they tend to come from the innermost disc $<1.5\,$AU for EJS.
Since the minimum semimajor axes of collided planetesimals are $\lesssim 1\,$AU for GT1.5AU, 
it is expected that these planetesimals mostly collided with embryos that eventually became planets.
These trends indicate that the inner planetesimals are well mixed in the GT model.
On the other hand, since planetesimals that ended up in planets appear to come most commonly 
from the innermost region in EJS, it is expected that planetesimals are less well-mixed in EJS compared to GT.
Indeed, as shown in Table~\ref{tab5}, most planetesimals stay in the initial region for EJS, 
while $\sim50\%$ of planetesimals from $1-2\,$AU and $\sim20\%$ from $2-3\,$AU arrive within $1\,$AU for GT models.

The distribution for CJS looks similar to GT and planetesimals from all over the inner disc collide with embryos.
However, the percentages of planetesimals that end up within $1\,$AU are much smaller than GT and $\sim26\%$ and $\sim10\%$ 
for planetesimals from $1-2\,$AU and $2-3\,$AU, respectively. 
Thus, it appears that the radial mixing of materials is most efficient in GT models, 
less efficient in CJS, and least efficient in EJS.
In Section~\ref{abundance_planets}, we will show that this leads to differences in compositions.  

The dominant fate of outer planetesimals for all the models is ejection out of the system.
However, percentages of outer planetesimals that reach the inner disc are about one order of magnitude lower 
in EJS compared to the other models for each region (see Table~\ref{tab5}).
Thus, the scattering of outer planetesimals appears to be much less efficient in EJS compared to the others. 
The scattering sources of outer planetesimals are different for GT and CJS models, and 
we will discuss this below.

\subsubsection{Collisional speed and the scattering of outer planetesimals}
The collisions between two bodies could lead to agglomeration or destruction, 
depending on the magnitude of the encounter speed relative to the escape speed.  
\cite{Marcus09} studied collisions between terrestrial-size planets ($1-10M_E$) 
via smoothed particle hydrodynamics simulations and showed that  
the erosive outcome is more common when the impact speed is larger than twice the escape speed 
($v_{\rm imp}/v_{\rm esc} > 2$), though the outcome strongly depends on the impact angle.

\begin{figure*}
\plotone{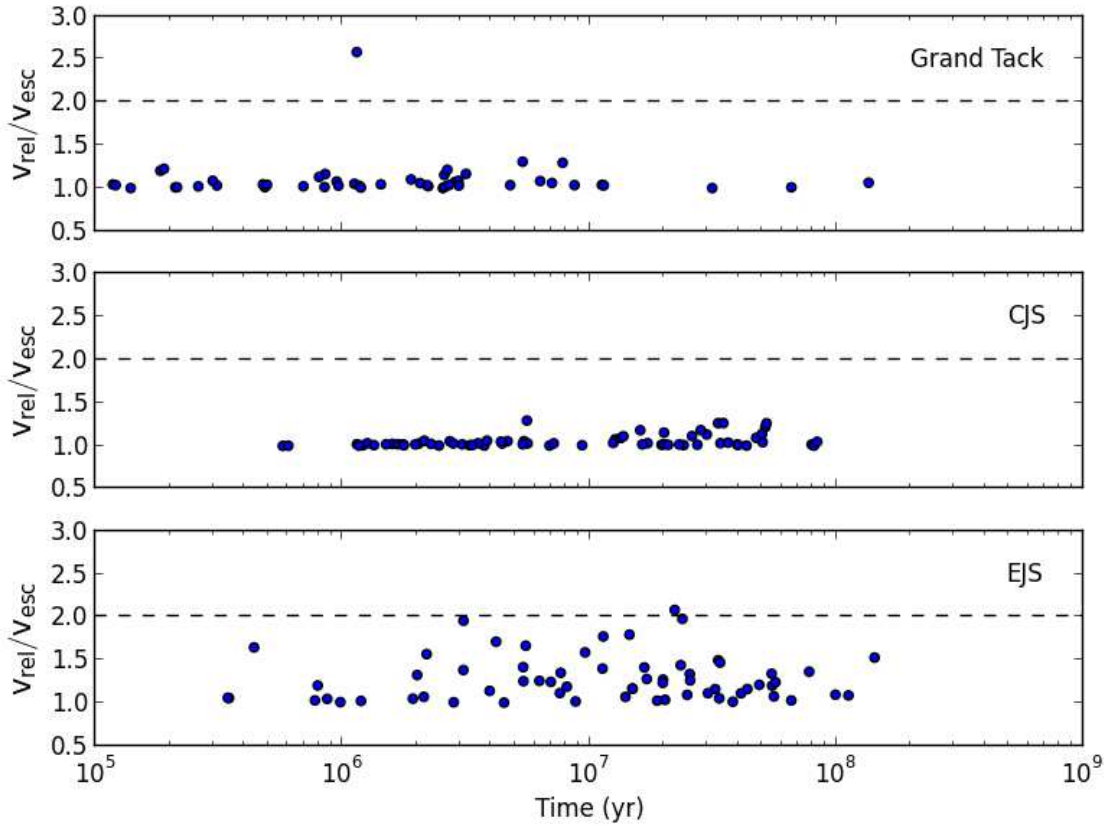}
\caption{The ratio of a relative encounter speed to a mutual escape speed as a function of time 
for embryo-embryo collisions for the GT1.5AU (top), CJS (middle), and EJS (bottom) models.
\label{vrel_vesc}}
\end{figure*}

\begin{figure*}
\plotone{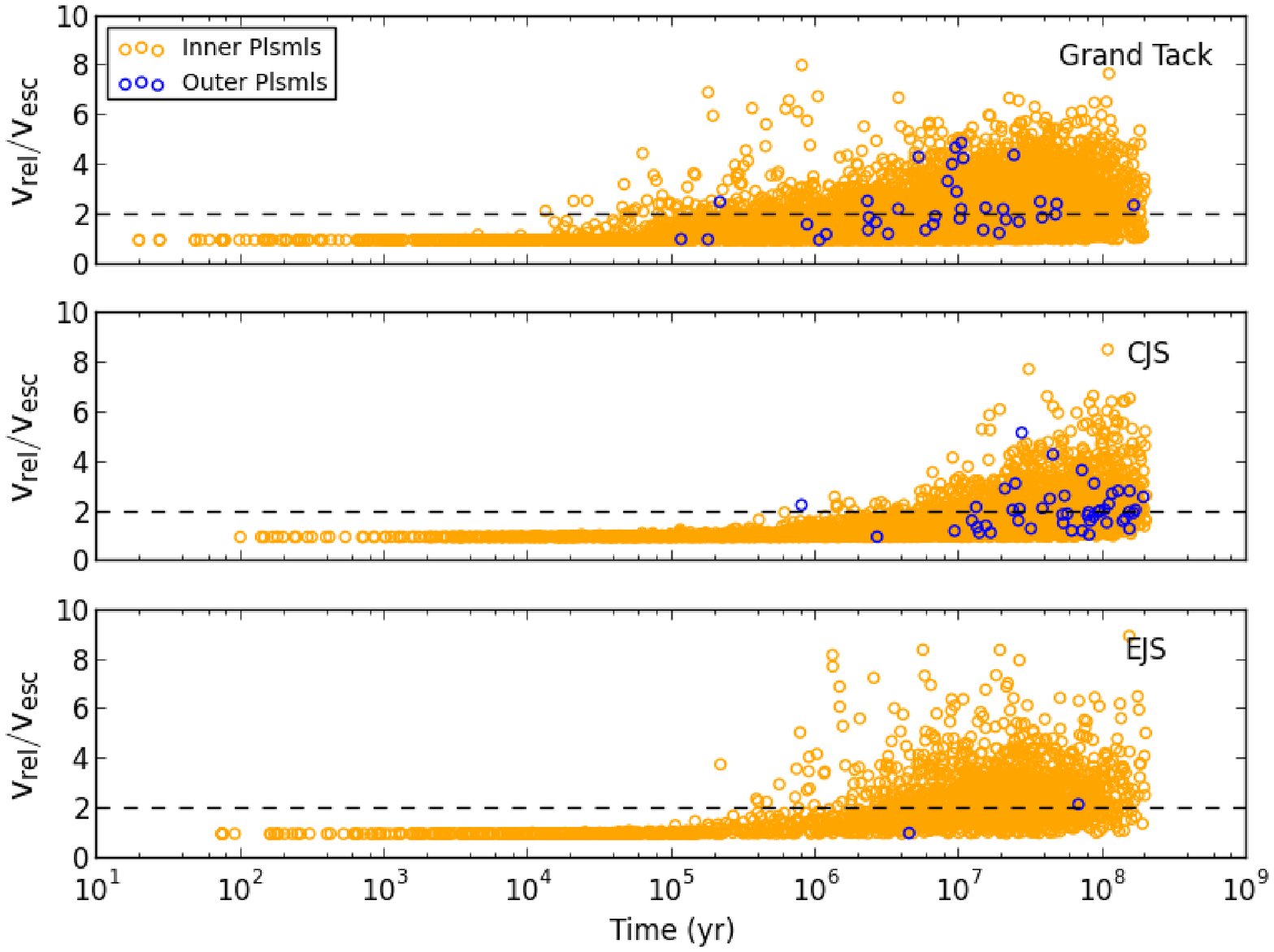}
\caption{The ratio of a relative encounter speed to a mutual escape speed as a function of time 
for embryo-planetesimal collisions for the GT1.5AU (top), CJS (middle), and EJS (bottom) models.
\label{vrel_vesc2}}
\end{figure*}

In Figures~\ref{vrel_vesc}, \ref{vrel_vesc2}, we show the ratios of relative encounter speeds 
of colliding bodies to mutual escape speeds as a function of simulation time for GT1.5AU, CJS, and EJS models.
Here, the mutual escape speed is defined as $v_{\rm esc} = \sqrt{2G(M_1 + M_2)/(R_1+R_2)}$.
For the GT1.5AU model, all embryo-embryo collisions but one have the speed ratio less than 1.5, which indicates the 
agglomeration rather than the erosion.  For embryo-planetesimal collisions, the ratio increases with time, and 
some collisions may be destructive.  There is no dependence of this ratio on the initial locations of planetesimals, since 
both inner and outer planetesimals (defined as planetesimals within and beyond 3 AU here) show similar distributions.

The outcomes for the CJS model are overall similar to the GT model. 
We find that all embryo-embryo collisions have the speed ratio less than 1.4, which indicates agglomeration.
The overall shape of the speed ratio for embryo-planetesimal collisions is similar to the GT case as well, with 
some of them being outer planetesimals.  This indicates that, even without a giant planet migration, the CJS case 
has an efficient inward scattering of outer planetesimals. 
The critical difference, however, is a timing of the scattering. 
For the GT case, the outer planetesimals are scattered inward and collide with embryos throughout the simulations, 
from $0.1-100\,$Myr.  On the other hand, for the CJS case, most inward scattering occurs after $10\,$Myr.
This difference arises because, in the CJS model, the scattering is done 
by growing terrestrial planets rather than by giant planets.
Indeed, Figure~\ref{scatter_aeim1} shows that more massive terrestrial planets tend to be 
at large orbital radii (i.e., near the asteroid belt) for the CJS model compared to the others.

The EJS model shows a similar trend for the embryo-embryo collision to the other cases, 
but the relative velocity varies more than GT or CJS models. 
This may be due to their slightly higher eccentricities and inclinations compared to the other models (see Table~\ref{tab3}).
The distribution of embryo-planetesimal collisions is similar to that of the GT and CJS models.  
However, in EJS model, there are almost no outer planetesimals that collide with embryos. 
The scattering is inefficient in EJS, because 
massive rocky planets do not exist in the asteroid belt region as in CJS, because 
they tend to be removed by secular effects (see Figures~\ref{scatter_aeim1} and \ref{fates_emb}).

An interesting difference we could see in Figure~\ref{vrel_vesc} is that the major impact starts and finishes earlier 
in GT models ($\sim0.1-10\,$Myr) compared to CJS and EJS models ($\sim1-100\,$Myr).
The early occurrence of giant impacts is due to the migration of giant planets.  
However, as we discuss below, the timing of the last giant impact 
depends on the initial mass ratios of planetesimals and embryos.

In summary, we find that the scattering of outer planetesimals is inefficient in EJS and 
efficient in CJS and GT. 
In GT, the giant planets scatter planetesimals in the outer disc to the inner disc, while in CJS, 
the terrestrial planets formed near the asteroid belt are responsible for the scattering.
As we show in Section~\ref{abundance_planets}, 
the difference in the scattering efficiency leads to the difference in abundances of highly volatile species.
\subsubsection{Moon-forming timescale}
\begin{figure*}
\plotone{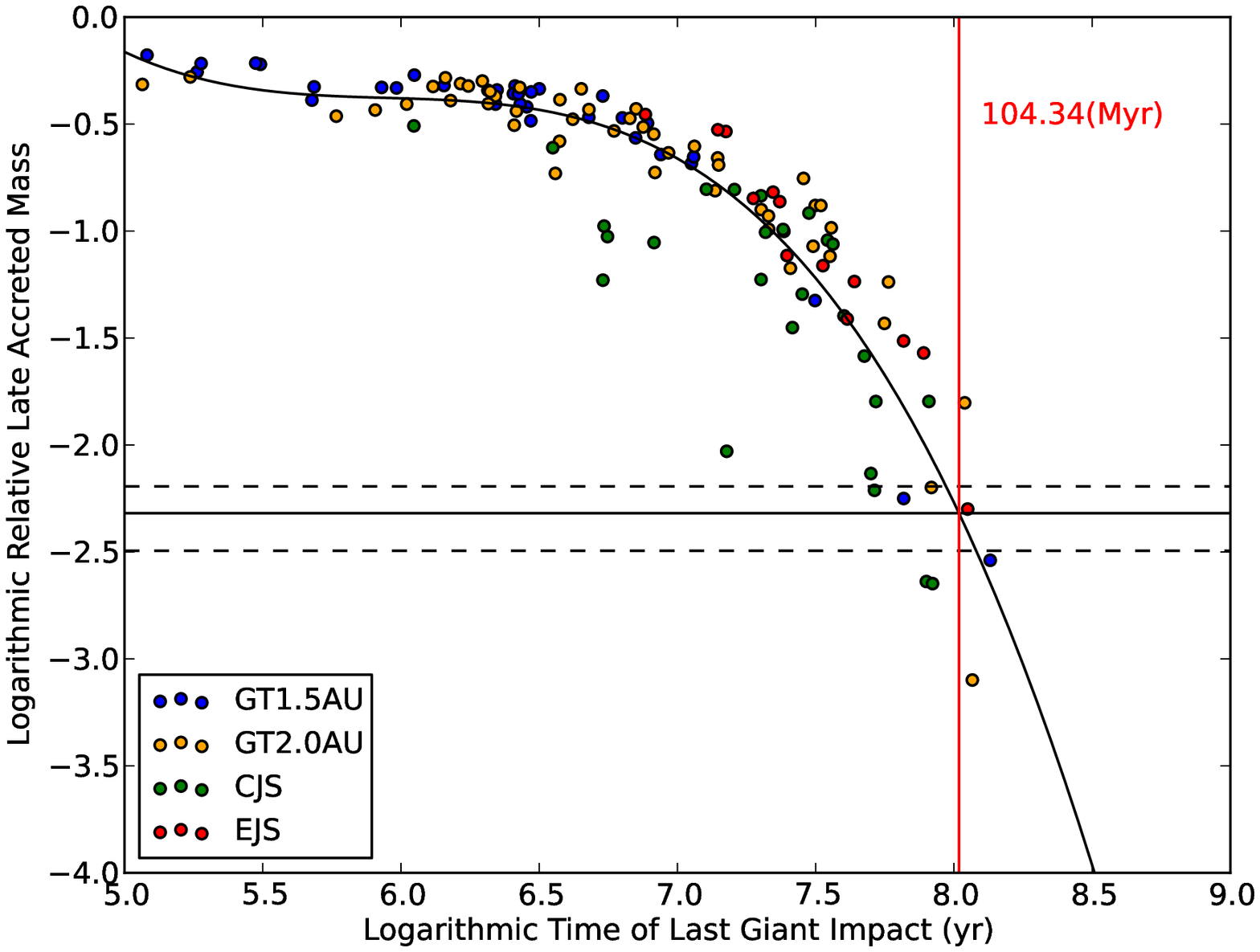}
\caption{The late accreted mass normalised by the planetary mass as a function of the last 
impact time between embryos.  The late accreted mass is the mass accreted onto an embryo after the last giant impact 
between embryos.
The horizontal solid and dashed lines are the best estimate and $1\sigma$ uncertainty 
of the late accreted mass estimated from the highly siderophile element abundances in the mantle: 
$4.8\pm1.6\times10^{-3}\,M_{\oplus}$ \citep{Jacobson14}. 
The intersection of the best estimate and the polynomial fit to the combined data ($104.34\,$Myr) is shown as a vertical red line. 
The intersection is $109.04\,$Myr for only the results from GT1.5AU and GT2.0AU. \label{lateaccr}}
\end{figure*}

For the completeness, we estimate the Moon-forming timescale in our simulations.
\cite{Jacobson14} calculated the late-accreted mass of a planet, which is the total mass of planetesimals accreted onto 
an embryo after the last giant impact between embryos, 
and found that the late-accreted mass decreases with the time of the last giant impact.  
They argued that the correlation could be used to estimate the timing of the Moon-forming event by 
comparing the estimated late-accreted mass with that of Earth, 
and proposed that the Moon-forming event could have happened long after $40\,$Myr and around $95\pm32\,$Myr.
Figure~\ref{lateaccr} is the corresponding figure for our simulations.
The estimated time is $\sim109\,$Myr for Grand Tack simulations, and is consistent with their estimate. 
Our simulations did not produce many planets which experienced the last impact at such late times, compared to \cite{Jacobson14}.
This is because our simulations assume the same total mass for embryos and planetesimals, 
while \cite{Jacobson14} includes various mass ratios.
As they showed in the supplementary information, 
the time of the last giant impact increases when the dynamical friction effect is 
reduced due to the smaller total mass in planetesimals compared to embryos.

For CJS and EJS simulations, we find a similar trend to GT runs.
By taking account of all of the simulations, the estimated time of the Moon-forming event 
is $\sim104\,$Myr.
The trend that CJS and EJS give slightly earlier time than GT 
is the opposite from \cite{Jacobson14}, since their corresponding simulations point to 
a comparable, but overall longer, last giant impact time for the Moon formation.  
It is unclear whether this is a statistically significant effect.
In \cite{Brasser16pre}, we will study various initial conditions for the GT model 
by taking account of type I migration and the size distribution of embryos to investigate the late accretion problem further.

\subsection{Elemental Abundances of Planets}\label{abundance_planets}
Overall, the dynamical results imply (1) the radial mixing of materials is most efficient in GT and 
less so in CJS and EJS models, and (2) icy outer planetesimals could be scattered into the inner disc region 
and thus may be able to contribute to the delivery of atmophile species to rocky planets 
in GT and CJS models, but not in EJS model.
In this subsection, we discuss how these differences affect the elemental abundances of planets.
All the simulated planets will be compared with the compositions of the Earth. 
This is partly because the compositions of Earth is the best known of all terrestrial planets, 
and the estimates for the other planets except for Mercury are similar to those of the Earth. 
The composition of the bulk Earth used in this study is taken from \cite{McDonough03} 
and is shown in Table~\ref{tab6}.
Below, we discuss the abundances of refractory and moderately volatile elements, water, 
and more volatile C and N elements separately.
\begin{deluxetable*}{lccccccc}
\tabletypesize{\small}
\tablecaption{Elemental Abundances used for HSC chemistry \label{tab6}}
\tablecolumns{8} 
\tablehead{ 
\colhead{Element Name} &  
\colhead{Earth} & \colhead{H} & \colhead{LL} & \colhead{CI} & \colhead{CM} & \colhead{EH} & \colhead{EH} }
\startdata              
$\rm H$ & $260\,$ppm & $-$ & $-$ & $2.0$ & $1.4$ & $-$ & $-$\\     
$\rm C$ & $730\,$ppm & $0.11$ & $0.12$ & $3.2$ & $2.2$ & $0.40$ & $0.36$ \\     
$\rm N$ & $25\,$ppm & $48\,$ppm & $70\,$ppm & $0.15$ & $0.152$ & $-$ & $-$ \\      
$\rm O$ & $29.7$ & $35.7$ & $40.0$ & $46.0$ & $43.2$ & $28.0$ & $31.0$ \\     
$\rm Na$ & $0.18$ & $0.64$ & $0.70$ & $0.49$ & $0.41$ & $0.68$ & $0.58$ \\    
$\rm Mg$ & $15.4$ & $14.0$ & $15.3$ & $9.7$ & $11.7$ & $10.6$ & $14.1$ \\    
$\rm Al$ & $1.59$ & $1.13$ & $1.19$ & $0.86$ & $1.18$ & $0.81$ & $1.05$ \\    
$\rm Si$ & $16.1$ & $16.9$ & $18.9$ & $10.5$ & $12.9$ & $16.7$ & $18.6$ \\    
$\rm P$ & $0.121$ & $0.108$ & $0.085$ & $0.102$ & $0.090$ & $0.200$ & $0.117$ \\    
$\rm S$ & $0.635$ & $2.0$ & $2.3$ & $5.9$ & $3.3$ & $5.8$ & $3.3$ \\    
$\rm Ca$ & $1.71$ & $1.25$ & $1.30$ & $0.92$ & $1.27$ & $0.85$ & $1.01$\\    
$\rm Ti$ & $813\,$ppm & $600\,$ppm & $620\,$ppm & $420\,$ppm & $580\,$ppm & $450\,$ppm & $580\,$ppm \\    
$\rm Cr$ & $0.466$ & $0.366$ & $0.374$ & $0.265$ & $0.305$ & $0.315$ & $0.305$ \\   
$\rm Fe$ & $31.9$ & $27.5$ & $18.5$ & $18.2$ & $21.0$ & $29.0$ & $22.0$ \\    
$\rm Ni$ & $1.82$ & $1.60$ & $1.02$ & $1.07$ & $1.20$ & $1.75$ & $1.30$  
\enddata
\tablecomments{Column 1: name of the element, and 
columns 2-8: elemental concentration (${\rm ppm}=1.e-4\%$) for Earth and various chondrites, 
where the values are in $\%$ unless stated otherwise.
The values for Earth are taken from Table~3 of \cite{McDonough03} as well as 
through a private communication with Bill McDonough in 2015. The values for chondrites are 
taken from \cite{Wasson88}. }
\end{deluxetable*}

\subsubsection{Refractory and moderately volatile elements}
\begin{figure*}
\plotone{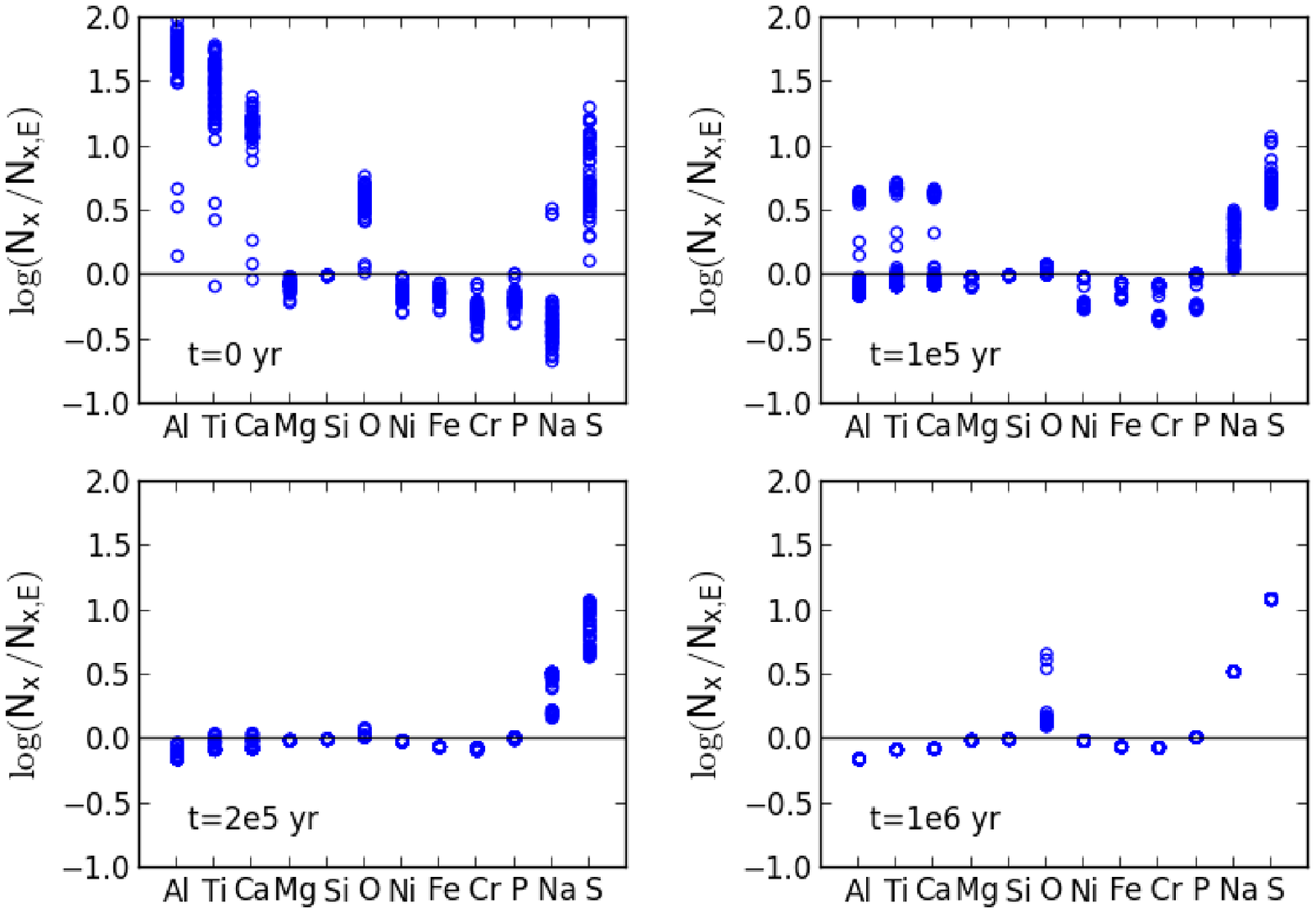}
\caption{The abundances of refractory and moderately volatile elements 
normalised by those of Earth for all survived embryos in GT1.5AU. 
The abundance ratio $N_X$ is taken with respect to Si, thus the ratio for Si is always 1.  
Different panels correspond to different initial disc ages.  \label{refractory_run0}}
\end{figure*}

Figure~\ref{refractory_run0} shows the elemental abundance comparisons for all the planets formed in GT1.5AU.
The abundances of refractory and moderately volatile elements are listed 
roughly in increasing order of volatility, 
and are normalised by the corresponding abundances of Earth for four different initial disc ages of 
$t_{\rm init, \, disc}=0$, $0.1$, $0.2$, and $1\,$Myr.
The abundance ratio for an element $X$ is defined as follows.
\begin{equation}
N_X= \frac{\rm wt.\% \, of \, X}{\rm wt.\% \, of \, Si}
\end{equation}
%
In our model, we find the best agreement for $t_{\rm init, \, disc}=0.2\,$Myr.
The abundances of refractory and moderately volatile elements agree with those of the Earth very well, 
except for some of the most volatile elements in this figure ($\rm N$a and $\rm S$).
This trend is consistent with previous works \citep[e.g.,][]{Bond10a,Elser12,Moriarty14}.
The agreement could be improved by taking account of a volatile loss, but such a study is 
beyond the scope of this paper.

\begin{figure*}
\plotone{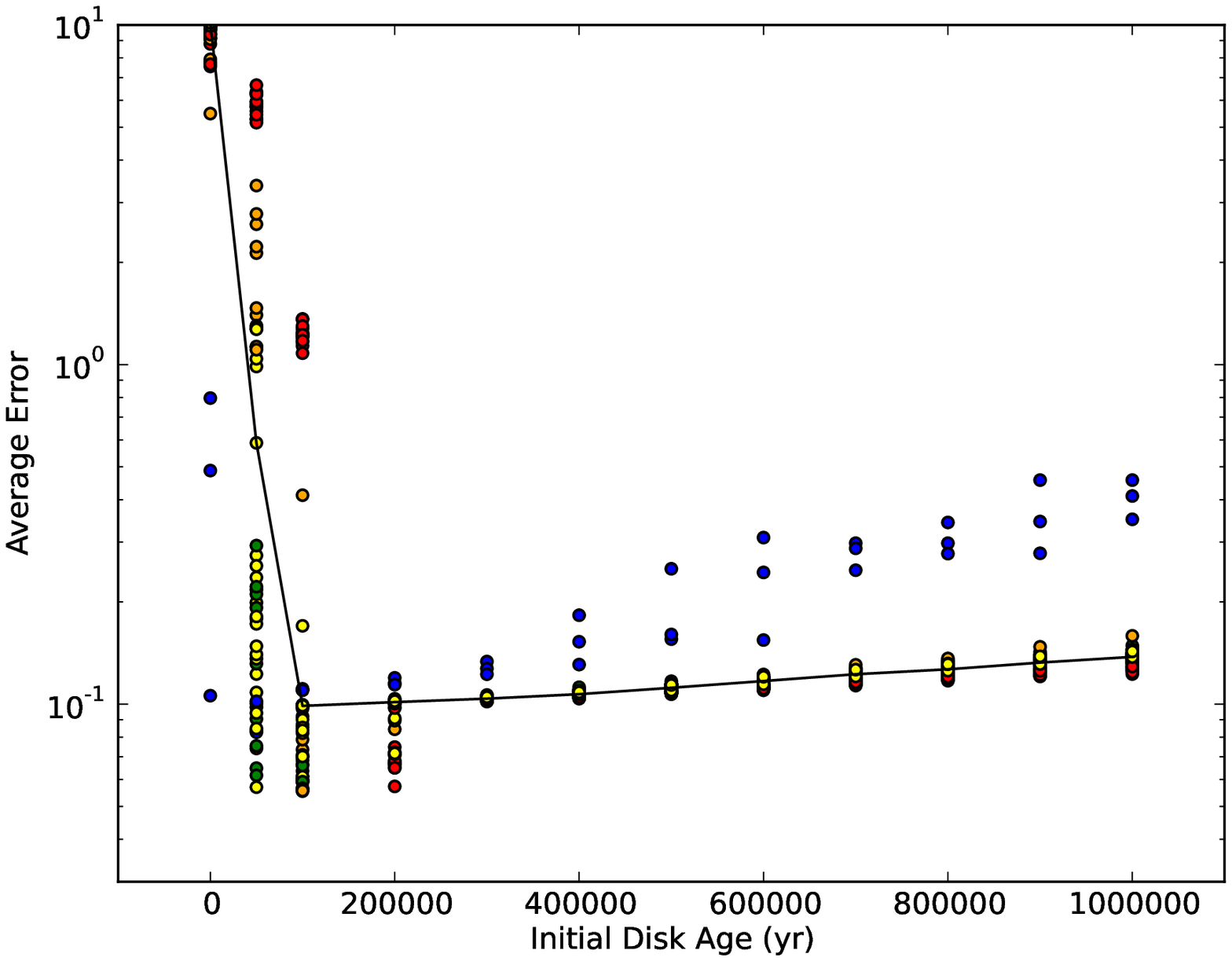}
\caption{The average errors of the abundance ratios with respect to Earth are shown for GT1.5AU runs 
with different initial disc ages.  All the elements listed in Figure~\ref{refractory_run0} 
are included except for $\rm Na$ and $\rm S$.  
The red, orange, yellow, green, and blue circles correspond to the initial semimajor axis 
of embryos being $a\leq 0.6\,$AU, $0.6\,{\rm AU} < a \leq 0.7\,$AU, $0.7\,{\rm AU} < a \leq 1.0\,$AU, 
$1.0\,{\rm AU} < a \leq 2.0\,$AU, and $2.0\,{\rm AU} < a \leq 3.0\,$AU, respectively. 
The solid line indicates the median error for all the planets at each initial disc age. 
\label{refractory}}
\end{figure*}
Figure~\ref{refractory} shows the overall agreement of the abundances of refractory and moderately volatile elements of 
planets from GT1.5AU with those of Earth 
for different initial disc ages (i.e., different initial compositions of embryos and planetesimals).
The error in the abundance ratio with respect to the Earth is calculated for each element except for $\rm Na$ and $\rm S$, and 
the average error was determined for each planet.  
The $\rm Na$ and $\rm S$ were not included here because these two tend to give the largest errors 
as in Figure~\ref{refractory_run0}.
The solid line indicates the median value of the calculated average errors for each $t_{\rm init, \, disc}$.
The overall average error becomes the smallest for the initial disc age of $0.2\,$Myr and 
increases on both sides.
For earlier times, the agreement is better for embryos which were initially further out, 
while for later times, the agreement is better for closer-in embryos.
The figure indicates that, if the initial disc was cold enough ($t_{\rm init, \, disc}\geq0.3\,$Myr), 
there is a little difference in abundances of refractory and moderately volatile elements for planets 
over a wide range ($\sim0.5-2.0\,AU$). 
The corresponding figures for GT2.0AU, CJS, and EJS models are all very similar to Figure~\ref{refractory}.
Therefore, we conclude that we cannot distinguish different formation models from refractory and moderately volatile elements alone. 

\subsubsection{Hydrous species}
\begin{figure*}
\plotone{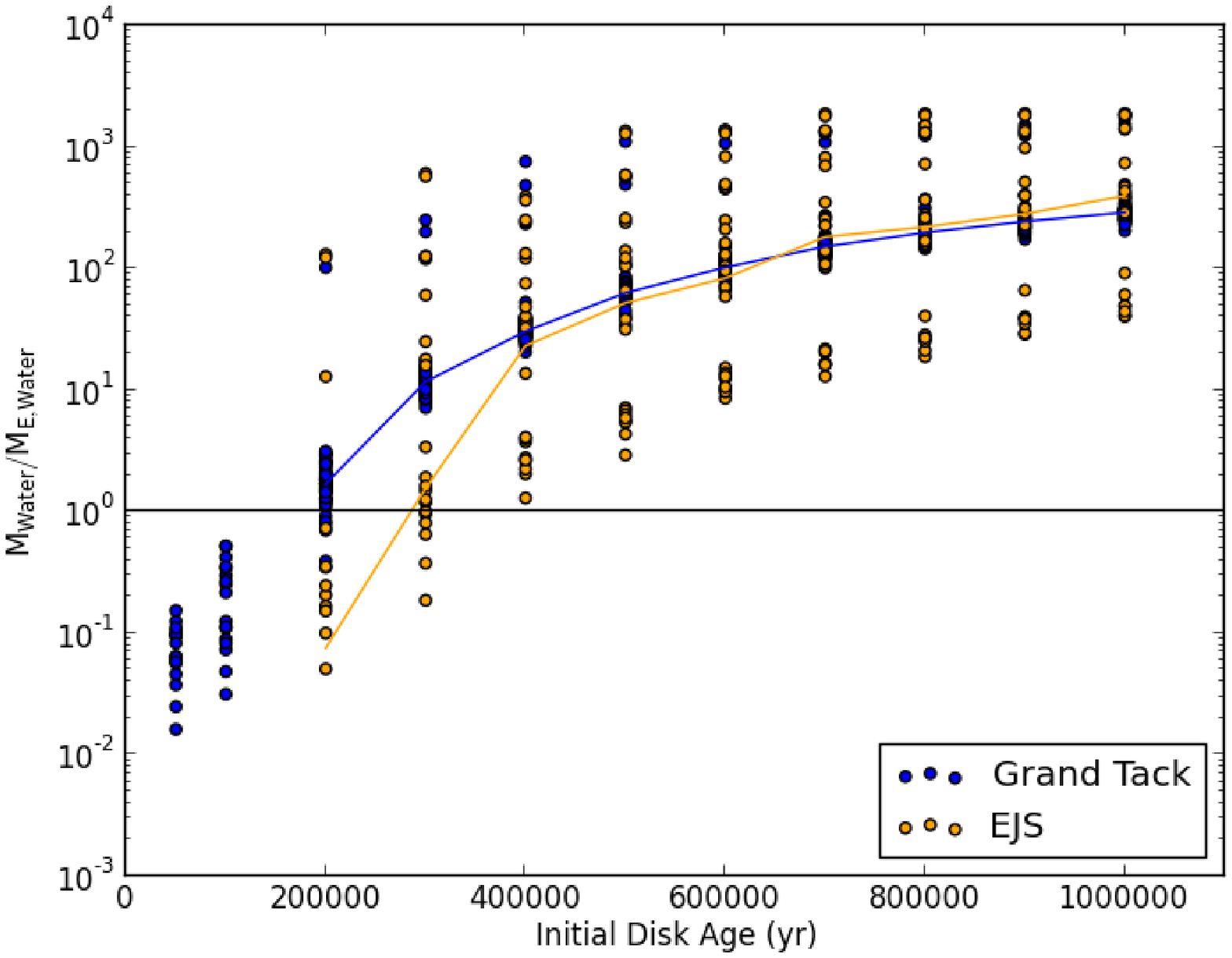}
\caption{A fraction of the water mass of a planet is normalised by that of the Earth for GT (blue) and EJS (orange) models, 
by assuming that the Earth's water mass is $2.34 \times 10^{-4}\,M_E$. 
The solid lines show the median values of the water content for these cases. 
For $t_{\rm init, \, disc} = 0.2\,$Myr and $0.3\,$Myr, planets in GT are accreting about one order of magnitude more water 
compared to those in EJS for the same initial disc age. \label{water}}
\end{figure*}

To investigate the abundance differences of more volatile species, 
we estimate the amount of water accreted onto a planet from hydrous species 
(water ice and serpentine $\rm Mg_3Si_2O_5(OH)_4$). Following \cite{Bond10a}, we assume that 
all of the hydrogen accreted as hydrous species is converted to water.  
The results are plotted in Figure~\ref{water} for GT1.5AU and EJS models.
The solid lines indicate the median values of a water mass fraction per planet normalised by that of the Earth.  
In both cases, the estimated amount of water increases with the initial disc age. 
The critical disc age to get water comparable to the Earth is 
$\sim 0.2\,$Myr and $\sim 0.3\,$Myr for GT and EJS, respectively. 
The results for the CJS model is similar to the GT model.
As seen in Figure~\ref{icelines}, the water iceline moves within $3\,$AU for $t_{\rm init, \, disc}\gtrsim 0.3\,$Myr while 
the serpentine iceline is slightly inward to water and moves into the inner disc for $t_{\rm init, \, disc}\gtrsim 0.2\,$Myr.
Since the hydrous species are included in some of the embryos and inner planetesimals beyond this time, 
the water delivery in EJS is largely explained by the inward migration of the icelines.
This is consistent with the lack of scattering of outer planetesimals into the inner disc in EJS model (see Figure~\ref{vrel_vesc2}).
Although the accreted fraction of water may seem fine for EJS, the model is not favourable to explain the water delivery.
This is because the requirement of the icelines being inside the inner disc region implies the 
formation of larger, icy cores, which are different from those expected for terrestrial planets.
%
%

On the other hand, for GT and CJS, outer planetesimals are clearly contributing to the delivery of water, since the 
water accretion becomes significant earlier than EJS.
Our results for EJS and CJS cases are in agreement with the results indicated in \cite{Bond10a}.
The comparable water accretion for GT and CJS indicates that 
both migration of giants and the presence of terrestrial planets in the asteroid belt region can help delivering water
via scattering of outer planetesimals.

\begin{figure*}
\plotone{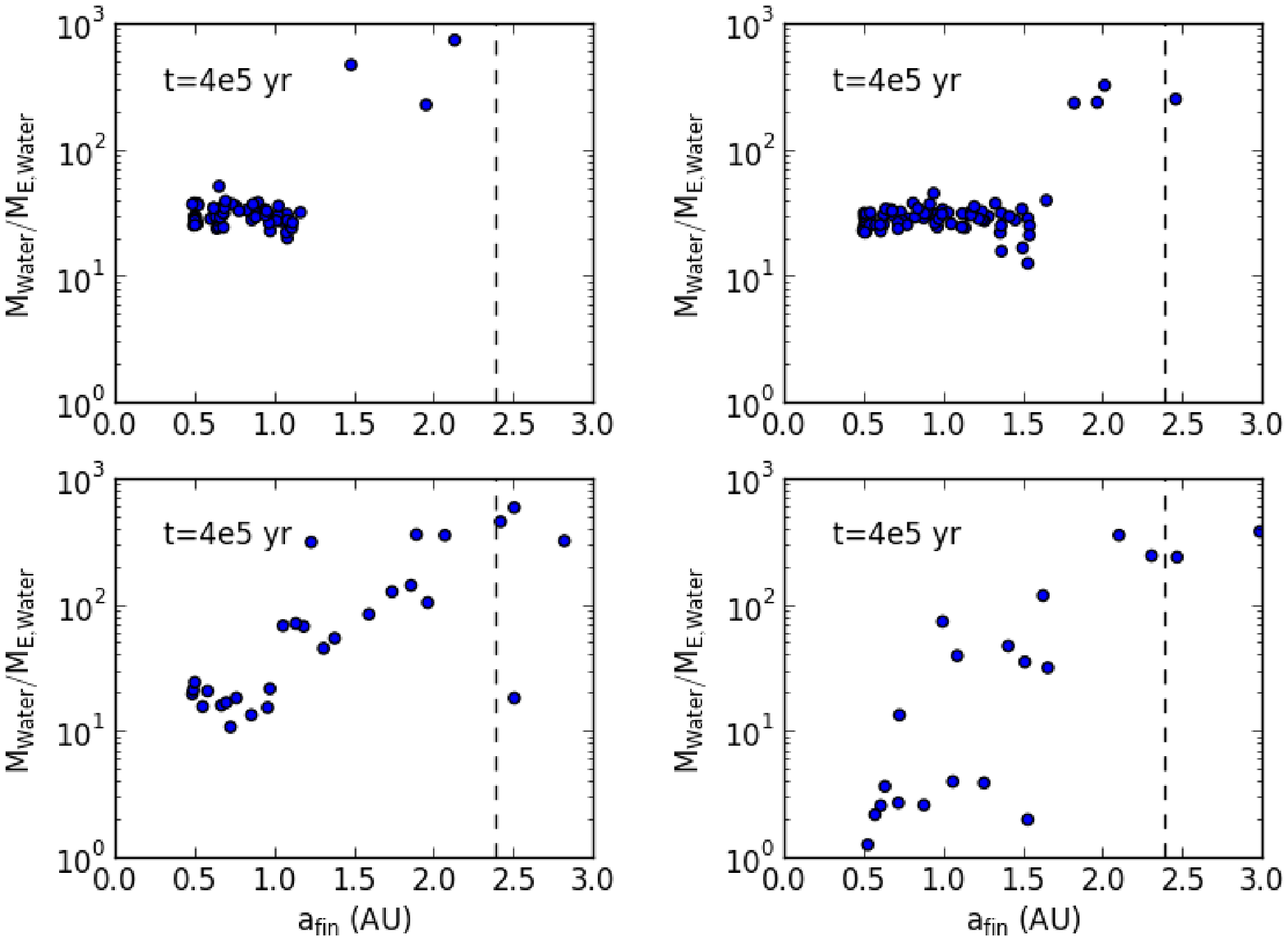}
\caption{A fraction of the water mass of a planet normalised by that of the Earth is plotted 
as a function of the final semimajor axis of a survived embryo 
for GT1.5AU (top left), GT2.0AU (top right), CJS (bottom left) and EJS (bottom right) models.
These are calculated for the Chamber's disc model at $t_{\rm init, \, disc}=0.4\,$Myr.
The vertical dashed lines indicate the location of the water iceline for this disc age ($\sim2.4\,$AU).
\label{water_comp}}
\end{figure*}
%
We also find that there is a correlation between water mass fraction and final semimajor axis.
Figure~\ref{water_comp} shows these results for four models by using the initial disc age of $t_{\rm init, \, disc}=0.4\,$Myr. 
The results are qualitatively similar for other initial disc ages.
For GT1.5AU and GT2.0AU cases, we find that the amount of accreted water is nearly constant independent 
of the initial semimajor axes of the survived embryos. 
This agrees with our expectation from Figure~\ref{fates_plsml} that the materials are well mixed in the GT model.
For EJS and CJS cases, on the other hand, the amount of water is weakly dependent on the initial semimajor axis 
of an embryo, and inner embryos tend to have a smaller amount of water.
For $t_{\rm init, \, disc}=0.4\,$Myr, the water iceline would be $\sim2.4\,$AU and thus embryos and planetesimals beyond this radius 
are expected to contain water initially.
The trend in water mass confirms that these water-rich materials did not reach the innermost disc as efficiently as GT models 
in CJS and EJS models, and that the materials are less well-mixed in these scenarios.

\subsubsection{Carbon and nitrogen}
\begin{figure*}
\plotone{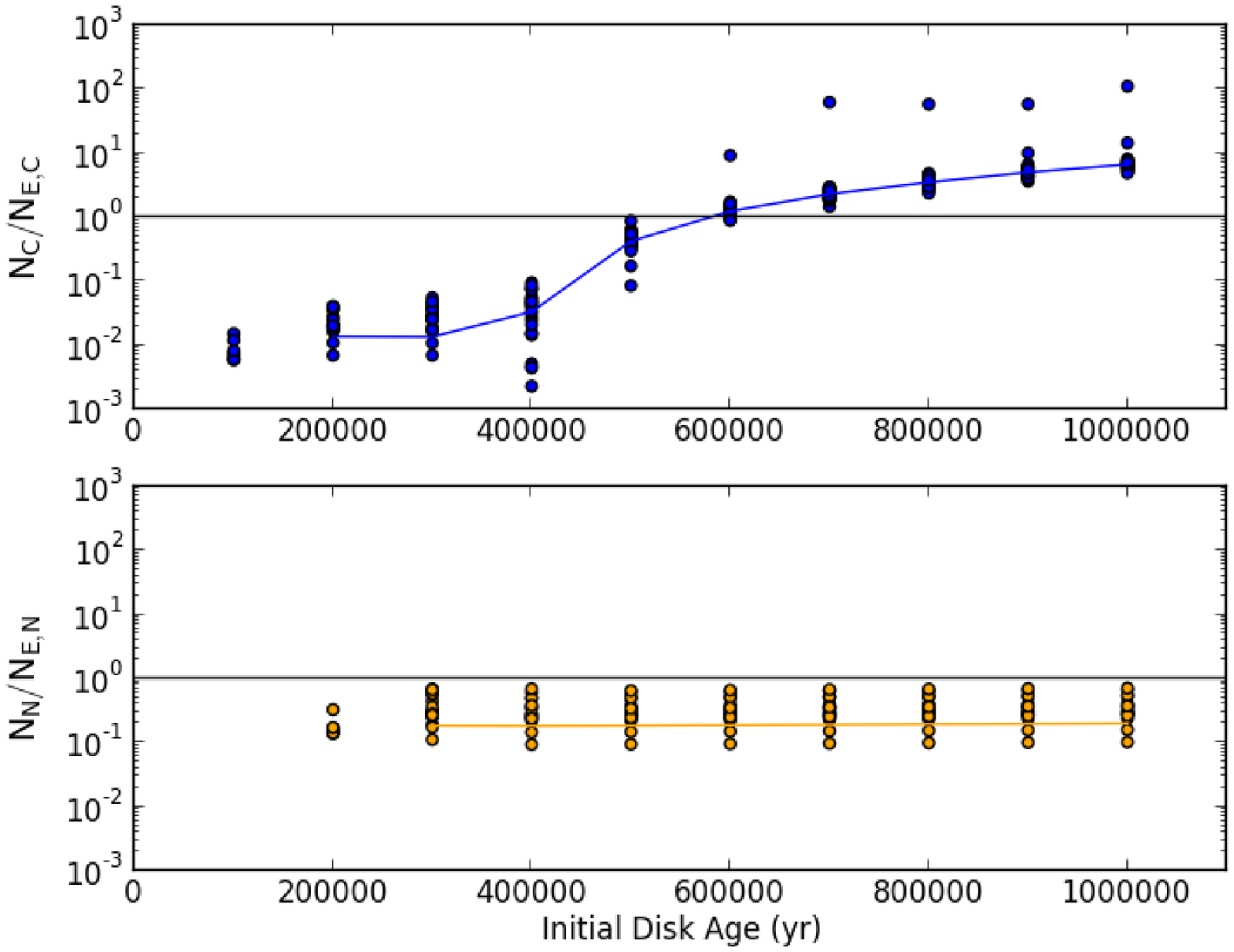}
\caption{The abundance ratio of $\rm C$ and $\rm N$ of a planet normalised by that of the Earth for GT1.5AU 
with an outer planetesimal disc. The solid lines show the median values of the $\rm C$ and $\rm N$ contents.  
The carbon comparable to Earth is delivered for the initial disc age of 
$0.6\,$Myr or beyond.  The nitrogen accretion rate is nearly independent of the disc's initial conditions.
\label{CNfraction1}}
\end{figure*}

We also check the abundances of even more volatile species --- carbon and nitrogen.
Figure~\ref{CNfraction1} shows how much $\rm C$ and $\rm N$ are delivered to terrestrial planets in GT1.5AU. 
For the case of carbon, we find that the abundance ratio comparable to Earth or larger is delivered 
for the initial disc age of $\geq0.6\,$Myr.
Since the iceline for $\rm CH_3OH$ enters within $3\,$AU for $t_{\rm init, \, disc}\geq0.4\,$Myr, 
this indicates that the $\rm C$ is incorporated into planets as a result of the iceline evolution.
For younger initial discs, the icelines of $\rm C$ species are further than $3\,$AU, 
and therefore the delivery of $\rm C$ should be solely done by outer planetesimals.
Since the abundances of $\rm C$ for $t_{\rm init, \, disc}<0.4\,$Myr are $\sim2$ orders of magnitude smaller than that of 
the Earth, we find that our default outer planetesimal disc is not massive enough to explain the abundance of $\rm C$ in the Earth.
%
%
\begin{figure*}
\plotone{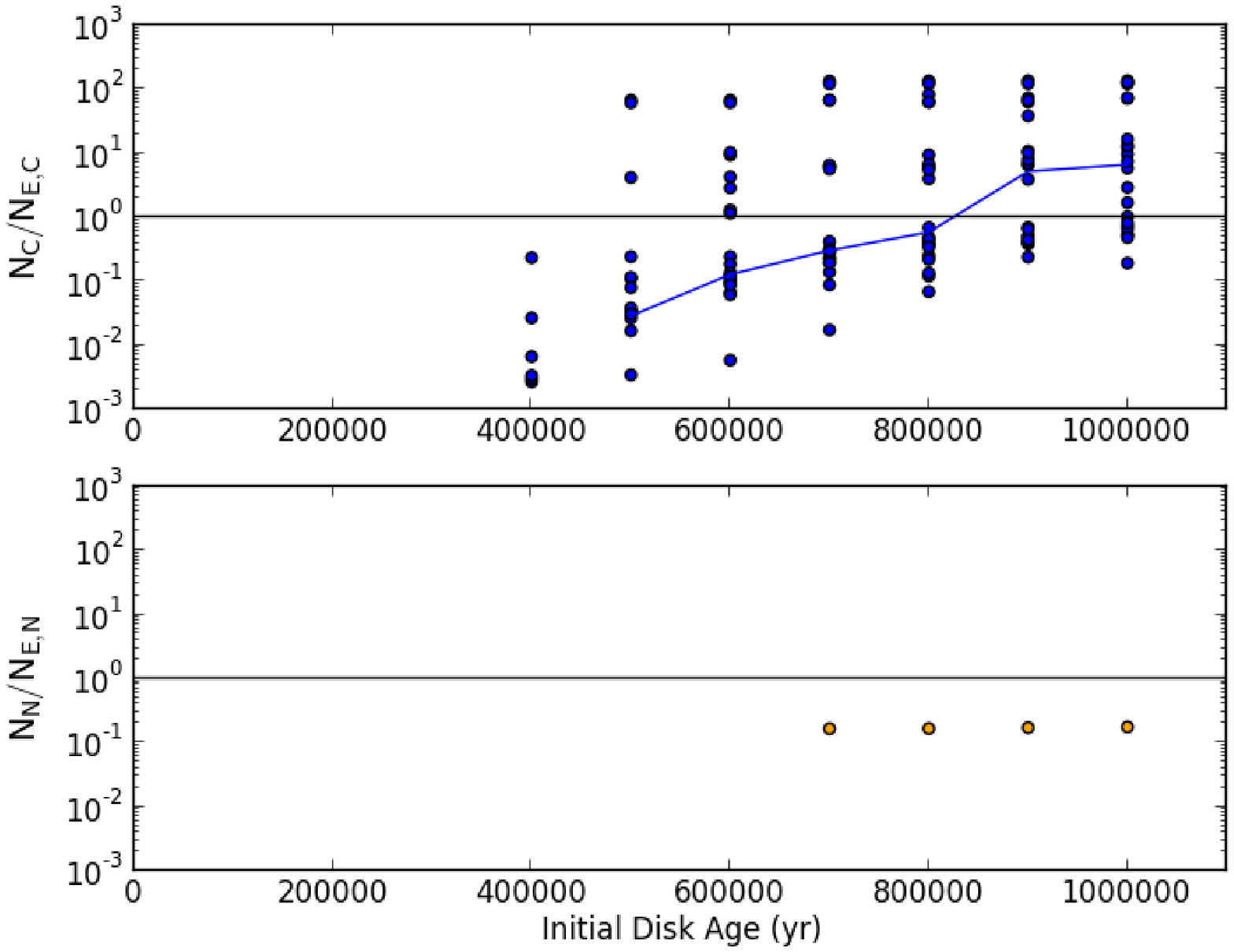}
\caption{The abundance ratio of $\rm C$ and $\rm N$ of a planet normalised by that of the Earth for EJS. 
The solid lines show the median values of the $\rm C$ and $\rm N$ contents.
The carbon comparable to Earth or larger is delivered for the initial disc age of 
$t_{\rm init, \, disc}\geq0.8\,$Myr. 
The median line for nitrogen does not show up because the accretion is inefficient.
\label{CNfraction2}}
\end{figure*}

On the other hand, the accreted amount of $\rm N$ is nearly independent of the initial disc age, indicating 
that all the $\rm N$ delivery is done due to the scattering of outer planetesimals.
This is consistent with the expectation from Figure~\ref{icelines}; 
since the $\rm NH_3$ iceline never enters within $3\,$AU, 
$\rm N$ is never incorporated initially within embryos and inner planetesimals.
However, the accreted amount of $\rm N$ is about an order of magnitude too low compared to the Earth in our model.

The results of the CJS model are similar to the GT results 
and is consistent with the expectation from Figure~\ref{vrel_vesc2}, where the efficient incorporation of 
outer planetesimals into embryos occurs for GT and CJS models.
The corresponding results for EJS are shown in Figure~\ref{CNfraction2}.
Since there is almost no scattering of outer planetesimals to the inner disc, the accretion of $\rm C$ occurs 
after the iceline enters $3$AU ($t_{\rm init, \, disc}\geq0.4\,$Myr) and the accretion of $\rm N$ occurs very rarely.
We confirm that the trend is similar to GT models {\it without} 
outer planetesimal discs (i.e., no scattering of these planetesimals into the inner disc region).
However, the variations of $\rm C$ and $\rm N$ abundances for the same disc age 
are much larger in the EJS model due to a similar radial dependence to Figure~\ref{water_comp}. 
Also, the median value of $\rm C$ becomes comparable to the Earth's amount for $t_{\rm init, \, disc}\geq0.6\,$Myr, 
while that occurs for $t_{\rm init, \, disc}\geq0.8\,$Myr in EJS.
This is also a consequence of the efficient radial mixing in GT models compared to EJS.

It may be possible to accrete larger amounts of $\rm C$ and $\rm N$ 
via more efficient gas drag on the outer planetesimals. 
For all the GT simulations shown so far, we have assumed a planetesimal size of $50\,$km for gas drag purposes.  
However, smaller planetesimals would migrate much faster, and thus icy materials 
could be incorporated into planets more efficiently.  
We assume three different planetesimal sizes ($1$, $10$, and $100\,$km) for gas drag, 
and run 8 simulations for each size until the gas disc dissipates.
We find that the mass percentages of outer planetesimals that arrive within $1.5\,$AU increase 
with decreasing planetesimal radii --- $2.9$, $1.9$, $0.75$, and $0.45\,$\% for $1$, $10$, $50$, and $100\,$km, respectively. 
Thus, if planetesimals are $1\,$km in size, about four times more icy materials 
could be delivered to rocky planets.   
However, as shown in Figure~\ref{CNfraction1}, the amount of $\rm C$ 
is lower by $\sim2$ orders of magnitude compared to Earth for default cases.  
Therefore, a more realistic size distribution of planetesimals is unlikely to resolve this issue.

%
To check whether it is possible to explain the delivery of $\rm C$ and $\rm N$ solely 
by the scattering of outer planetesimals, we assume that 
the outer planetesimal mass is a factor of $\sim 20$ larger and is comparable to the inner one ($2.3\times 10^{-3}\,M_E$). 
This means that the outer planetesimal disc mass changes from $\sim0.06\,M_E$ to $\sim1.2\,M_E$. 
In this case, the accreted amounts of $\rm C$ and $\rm N$ increase and become comparable to those of Earth, nearly independent of 
the initial disc age. 
This change in the mass of outer planetesimals also affects the amount of water accreted on planets. 
However, the effect is largely limited to the initial disc's age below $0.2\,$Myr, where 
the accreted amount of water becomes comparable to the amount of Earth's water.
The accreted amounts of $\rm C$ and $\rm N$ show a similar dependence on a semimajor axis to the case of water in Figure~\ref{water_comp}. 

The abundance of atmophile elements could potentially distinguish GT, CJS, and EJS models.
For EJS model, the delivery of atmophile species via outer planetesimals is inefficient. 
Therefore, the main building blocks of terrestrial planets (i.e., embryos and inner planetesimals) would 
need to include these species due to the iceline evolution. 
However, such a scenario could lead to the larger errors for abundances of more refractory species 
as well as the existence of larger, icy planetary embryos in the inner disc.
For GT and CJS models, the delivery of atmophile species could be aided by outer planetesimals scattered 
into planet formation region.
Distinguishing these two models is harder, but the timing of accretion may be later 
for CJS model than GT (see Figure~\ref{vrel_vesc2}) and the abundance may be dependent on the distance from the Sun 
for CJS and independent for GT (see Figure~\ref{water_comp}).

In summary, to satisfy the constraints from both refractory elements and the amount of water, 
the initial disc age of $\sim0.2\,$Myr is most preferable. 
This in turn implies that, if the $\rm C$ and $\rm N$ are incorporated into rocky planets during their formation, 
the scattering of outer planetesimals, rather than the iceline evolution, 
would be responsible for the delivery of these species.

\subsection{Elemental Abundances of Planetesimals}
\begin{figure*}
\plotone{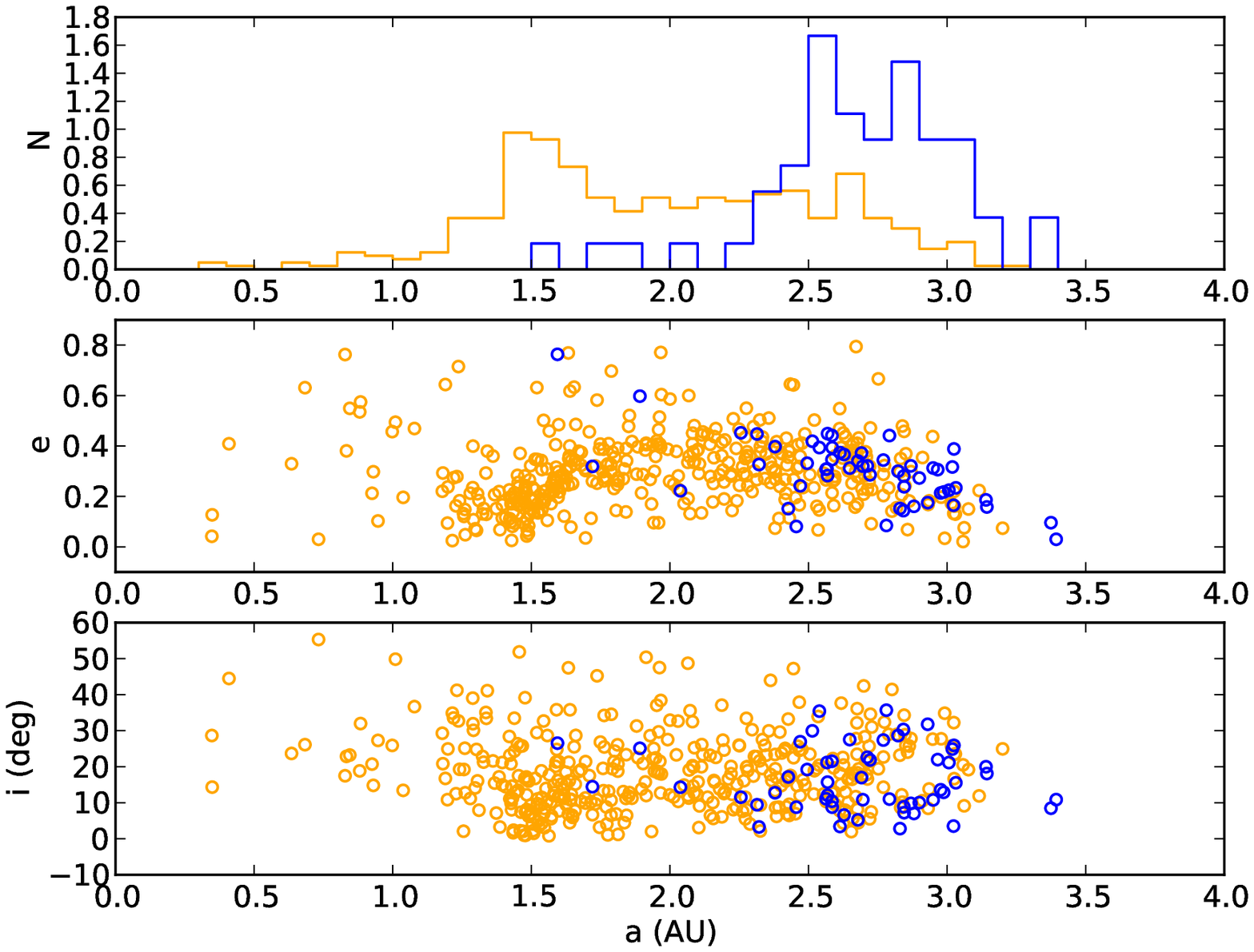}
\caption{Top panel shows the normalised orbital distribution of planetesimals which are initially in the inner disc (orange) 
and the outer disc (blue).  Middle and bottom panels show eccentricity and inclination values of these planetesimals 
as a function of a semimajor axis. \label{plsmls_orbits}}
\end{figure*}
Finally, we discuss the distributions and compositions of survived planetesimals.
Figure~\ref{plsmls_orbits} shows the final distribution of planetesimals for the GT1.5AU case.  
Most survived planetesimals are near the asteroid belt location ($\sim1.8-3.5\,$AU), 
and inner planetesimals (orange), which are initially within $3\,$AU, are located closer to the star compared to 
outer planetesimals scattered into the inner disc (blue).  
The average eccentricity is $0.288$ and the average inclination is $17.7^{\circ}$, and thus are 
more excited compared to the observed asteroids.
These trends are mostly consistent with \cite{Walsh11}.

\begin{figure*}
\plotone{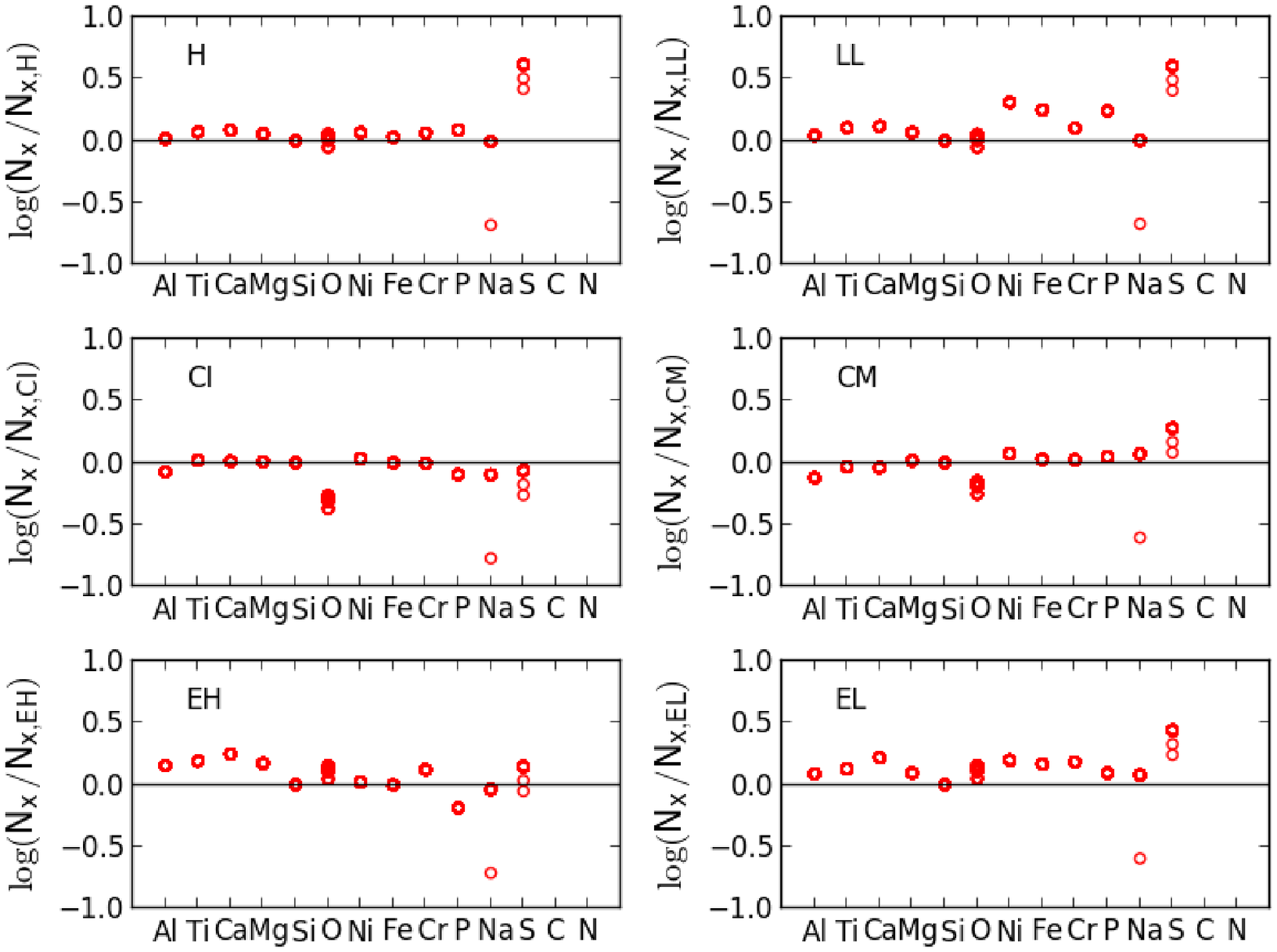}
\caption{The abundances of elements in inner planetesimals normalised by those of H (top left panels), LL (top right), CI (middle left), 
CM (middle right), EH (bottom left), and EL (bottom right) chondrites.  The initial disc age of $0.2\,$Myr is assumed here.   
Similar to Figure~\ref{refractory_run0}, the abundance ratio $N_X$ is taken with respect to Si, thus the ratio for Si is always 1.  
\label{abundance_plsmls}}
\end{figure*}
\begin{figure*}
\plotone{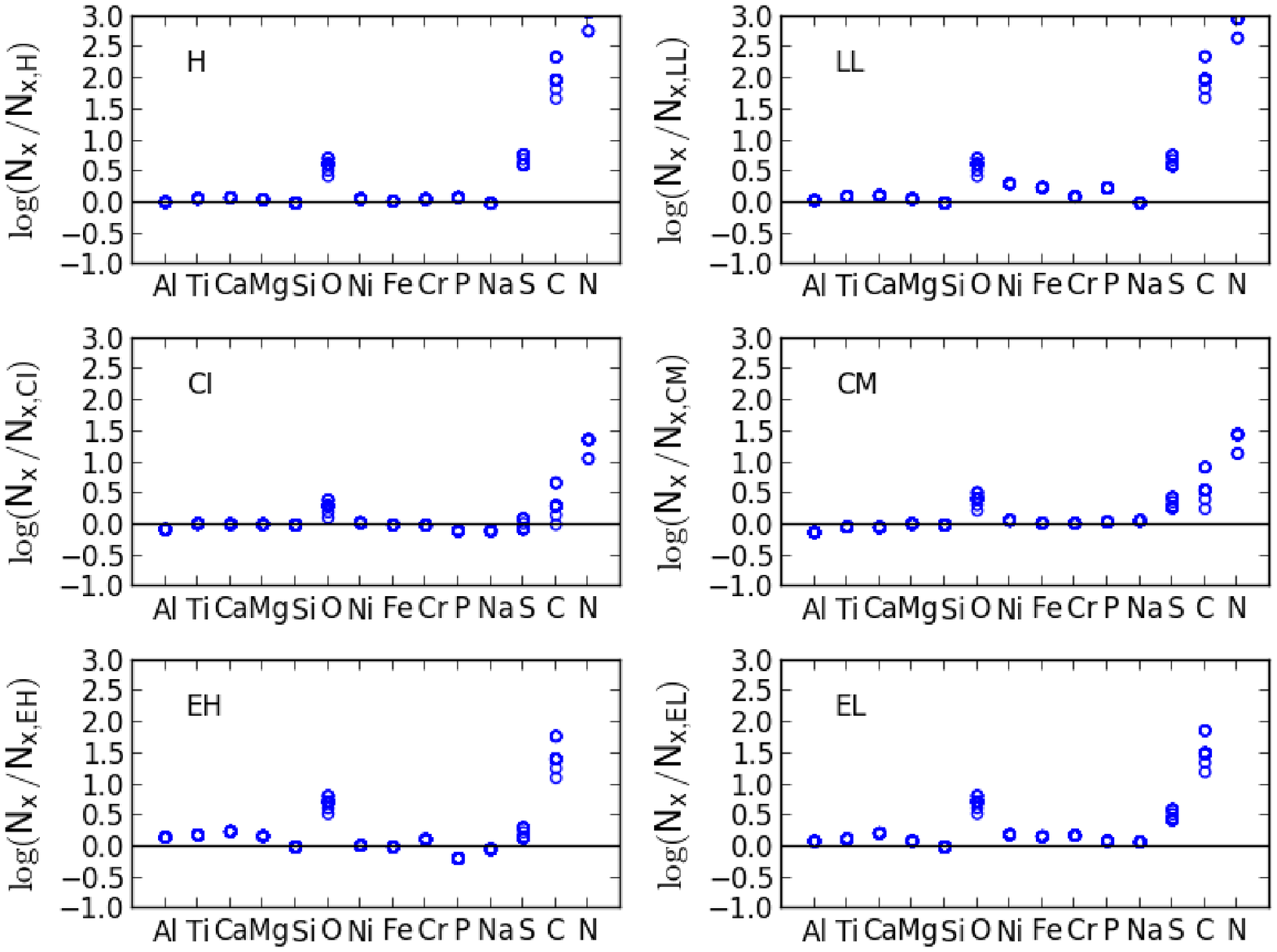}
\caption{The abundances of elements in outer planetesimals normalised by those of H (top left panels), LL (top right), CI (middle left), 
CM (middle right), EH (bottom left), and EL (bottom right) chondrites.  The initial disc age of $0.2\,$Myr is assumed here.   
Similar to Figure~\ref{refractory_run0}, the abundance ratio $N_X$ is taken with respect to Si, thus the ratio for Si is always 1.  
\label{abundance_plsmls2}}
\end{figure*}
To compare the compositions of survived planetesimals with those of chondrites, 
we choose planetesimals with semimajor axes within $1.8-3.5\,$AU, eccentricities $\leq 0.5$, 
and inclinations $\leq 40^{\circ}$ at the end of the simulations, which cover roughly 
the same ranges of orbital parameters to asteroids.
Figures~\ref{abundance_plsmls} and \ref{abundance_plsmls2} show the elemental abundances of 
the survived inner and outer planetesimals, respectively, 
compared to H, LL, CI, CM, EH, and EL chondrites.   
Since the inner planetesimals are more rocky than the outer ones, it is expected that 
they have more similar properties to S-type chondrites (H and LL) 
rather than C-type (CI and CM) or E-type (EH and EL) chondrites.
These figures show that inner planetesimals have a good agreement with H chondrites, 
except for $\rm S$, $\rm C$, and $\rm N$. 
Note that, since we ignore the interactions of planetesimals, 
their compositions do not change during the simulations due to collisions.
Also, for $t_{\rm init, \, disc}=0.2\,$Myr, the C and N icelines are not within the inner disc.
The abundances of $\rm O$ of inner planetesimals tend to be depleted compared to CI and CM. 
Compared to LL, EH, and EL chondrites, survived inner planetesimals tend to have overestimated refractory elements.

On the other hand, the outer planetesimals show a good agreement with CI chondrites except for $\rm N$, while 
the abundances of $\rm O$, $\rm S$, $\rm C$, and $\rm N$ tend to be enhanced compared to H, LL, EH, and EL chondrites. 
Therefore, our results show that inner planetesimals are more consistent with S-type chondrites while outer planetesimals 
are more consistent with C-type chondrites. 
These agree with the expectation from the Grand Tack scenario \citep{Walsh11}.


%
\section{Discussions and Conclusions}\label{discussions}
%
%
%
We have studied how the planetary compositions change depending on the dynamical effects of giant planets 
by testing three popular setups of terrestrial planet formation in the Solar System (GT, CJS, and EJS). 
The abundances of refractory and moderately volatile elements in terrestrial planets are nearly independent of 
orbital radii for $t_{\rm init, \, disc}\geq0.2\,$Myr for all the models 
(see Figure~\ref{refractory} and the discussion in text). 
Therefore, we cannot distinguish different formation models from these elements. 
The mass fractions of water and other more volatile species (e.g., $\rm C$ and $\rm N$), on the other hand, 
can be used to distinguish different models.

We have found that there are two key mechanisms to distinguish different models 
--- the radial mixing of the inner planetesimal disc and the scattering of outer, icy planetesimals into the inner disc.
The former tends to homogenise the planetary compositions over the inner disc, 
while the latter helps deliver the atmophile species, which only condense in the cold region of a disc, to 
terrestrial planets.
The dynamical simulations show that the mixing is efficient in GT models but less so in CJS and least efficient in 
EJS (see Table~\ref{tab5}).
On the other hand, the scattering of outer planetesimals is efficient both in GT and CJS models, while that is 
inefficient in EJS (see Figure~\ref{vrel_vesc2}).
These effects together lead to the following trends for each formation model.

In the classical formation model (EJS), neither the scattering nor the mixing are efficient. 
This makes it very difficult to deliver atmophile species to terrestrial planets.
Thus, if terrestrial planets are formed in the presence of giant planets with moderately eccentric orbits, 
we expect that such planets would be dry.
It is still possible for these planets to receive atmophile species as the disc cools and 
the icelines move inward. 
However, in such a case, the formed planets may not be terrestrial-size but rather like mini-Neptunes, because 
their cores may be born beyond the iceline and thus grow larger than typical rocky planets. 
Due to the poor radial mixing, planets in such a system may have a larger fraction of atmophile species 
for larger orbital radii.

In the GT model, both the scattering and the mixing are efficient.
The atmophile elements can be efficiently delivered to terrestrial planets, and the mass fractions of 
these elements would be at least initially nearly independent of orbital radii.
For the Solar System case, the GT model can reproduce the overall abundances of most 
refractory and moderately volatile elements, the amount of water, and 
the abundances of $\rm C$ and $\rm N$ consistently, 
if we assume a massive enough outer planetesimal disc with a total mass $\gtrsim1.2M_E$.
This disc condition further produces the asteroid analogues which have comparable compositions 
to S-type and C-type chondrites, 
where the S-type analogues tend to locate interior to the C-type ones, as shown in \cite{Walsh11}.
%

The CJS model is a ``hybrid'' of these two models, where the scattering is efficient but 
the mixing is not as efficient as the GT model.
The model would allow terrestrial planets to form much further than EJS and GT models, 
and the growing rocky planets on such orbits (rather than giant planets) could 
efficiently scatter the icy planetesimals into the inner disc.
We calculated Tisserand's parameter with respect to Jupiter in our simulations, and 
found that the dynamics of planetesimals within $\sim3\,$AU are not controlled by Jupiter.
The atomophile elements would be delivered to terrestrial planets, but the mass fractions 
of these elements may increase with orbital radii, different from the GT model (see Figure~\ref{water_comp}).
The CJS model is inappropriate to describe the formation path of the Solar System, because 
the delivery of water and other atmophile species relies on the presence of terrestrial planets 
in the asteroid belt region that do not exist now. 
The model has a prediction for some of the closely-packed extrasolar planetary systems that 
the delivery of atmophile species to rocky planets may be possible 
even in the systems which have no giant planets or those without strong perturbations from them. 
A recent study suggested that inner planets of systems with no giant planets are in fact more water-rich 
because the radial drift of icy materials would not be stopped by giants \citep{Morbidelli15ap}.

Our model relies on a critical assumption that the compositions of building blocks of terrestrial planets 
are determined by the condensation sequence.
However, the compositions of protostellar discs would be altered by many other mechanisms 
such as transient heating events and transport of chemical species, and also 
affected by the presence of grains inherited from the molecular cloud core. 
Moreover, the loss of volatiles during the impacts \citep[e.g.,][]{Okeefe82}, or the erosion of embryos 
and planetesimals \citep{Carter15ap} could also affect the final compositions of planets. 
The former may help explaining the overabundance of $\rm Na$ and $\rm S$, which are common in this type of studies 
\citep[our work and e.g.,][]{Bond10a,Elser12,Moriarty14}. However, that would also affect the abundance of 
more volatile species such as $\rm C$ and $\rm N$. 
The inclusion of these effects will be important for the future study.

In this work, we have not studied the compositions of gas giants. 
It has been proposed that the C/O ratio of a planetary atmosphere is important in 
characterising the compositions of an exoplanet \citep{Madhusudhan12}, 
and potentially distinguishing different formation models of hot Jupiters \citep{Madhusudhan14}.
Recent studies highlighted the importance of the locations of icelines \citep{Oberg11,AliDib14} as well as 
those of disc evolution and planet migration in determining the C/O ratio \citep{Moriarty14,Thiabaud15a}. 
The future study of planet formation will need to incorporate chemical processes as well as dynamical evolution 
to better understand the Solar System as well as extrasolar planetary systems.
%
\acknowledgements{We thank Kevin Walsh for providing us with SyMBA including a Grand Tack model.   
We thank an anonymous referee for the careful reading and comments.  
SM thanks Prof Bill McDonough, Prof Qing-Zhu Yin, and Prof Yuri Aikawa for useful discussions. 
SM thanks supports provided by the Kavli Institute for Theoretical Physics in the USA and 
the Earth-Life Science Institute in Japan, where some parts of this work were done, 
and a support from the Northern Research Partnership. 
RB is supported by the Astrobiology Centre Project of National Institutes of 
Natural Sciences (NINS) (Grant Number AB271017).}
\bibliographystyle{apj}
\bibliography{REF}

\end{document}